\pdfoutput=1
\newcommand{\PaperPath}{}   

\documentclass[preprint,12pt]{colt2024} 

\jmlrvolume{} \jmlryear{} \jmlrpages{} \jmlrproceedings{PMLR}{}

\newcommand{\qed}{\blacksquare}

\usepackage{nicefrac}
\newcommand{\xhdr}[1]{\vspace{1mm} \noindent{\bf #1}}

\numberwithin{equation}{section}
\usepackage[labelfont=bf,format=plain,justification=raggedright,singlelinecheck=false]{caption}

\usepackage[suppress]{color-edits}  
\addauthor{as}{red}      
\addauthor{bl}{blue}     
\addauthor{mz}{orange}   
\addauthor{sp}{green}    

\newenvironment{OneLiners}[1][\ensuremath{\bullet}]
    {\begin{list}
        {#1}
        {\setlength{\itemsep}{0pt}
         \setlength{\parsep}{0pt}
         \setlength{\topsep }{0pt}}}
    {\end{list}}

\newcommand{\inner}[1]{ \left\langle {#1} \right\rangle }

\newcommand{\order}{{\mathcal{O}}}
\newcommand{\calT}{{\mathcal{T}}}
\newcommand{\overlinev}{{\overline{v}}}

\newcommand{\quality}{c} 
\newcommand{\distrC}{F_{\mathtt{c}}} 

\newcommand{\bmax}{b_{\max}} 
\newcommand{\RegPow}{\alpha} 

\usepackage[capitalize,noabbrev]{cleveref}

\newtheorem{assumption}[theorem]{Assumption}
\newtheorem{observation}[theorem]{Observation}
\newtheorem{claim}[theorem]{Claim}

\title[Autobidders with Budget and ROI Constraints]%
{Autobidders with Budget and ROI Constraints:\\ Efficiency, Regret, and Pacing Dynamics}
\usepackage{times}

\coltauthor{%
 \Name{Brendan Lucier} \Email{brlucier@microsoft.com}\\
 \addr Microsoft Research, Cambridge
 \AND
 \Name{Sarath Pattathil} \Email{sarathp@mit.edu}\\
 \addr Massachusetts Institute of Technology
 \AND
 \Name{Aleksandrs Slivkins} \Email{slivkins@microsoft.com}\\
 \addr Microsoft Research, New York
 \AND
 \Name{Mengxiao Zhang} \Email{mengxiao-zhang@uiowa.edu}\\
 \addr University of Iowa
}
\usepackage{prettyref}
\usepackage{hyperref}
\newcommand{\pref}[1]{\prettyref{#1}}

\newcommand{\savehyperref}[2]{\texorpdfstring{\hyperref[#1]{#2}}{#2}}
\newrefformat{eq}{\savehyperref{#1}{Eq. \textup{(\ref*{#1})}}}
\newrefformat{eqn}{\savehyperref{#1}{Eq. \textup{(\ref*{#1})}}}
\newrefformat{tab}{\savehyperref{#1}{Table~~\ref*{#1}}}
\newrefformat{lem}{\savehyperref{#1}{Lemma~\ref*{#1}}}
\newrefformat{lemma}{\savehyperref{#1}{Lemma~\ref*{#1}}}
\newrefformat{def}{\savehyperref{#1}{Defn~\ref*{#1}}}
\newrefformat{line}{\savehyperref{#1}{Line~\ref*{#1}}}
\newrefformat{foot}{\savehyperref{#1}{Footnote~\ref*{#1}}}
\newrefformat{thm}{\savehyperref{#1}{Theorem~\ref*{#1}}}
\newrefformat{corr}{\savehyperref{#1}{Corollary~\ref*{#1}}}
\newrefformat{cor}{\savehyperref{#1}{Corollary~\ref*{#1}}}
\newrefformat{sec}{\savehyperref{#1}{Section~\ref*{#1}}}
\newrefformat{app}{\savehyperref{#1}{Appendix~\ref*{#1}}}
\newrefformat{assum}{\savehyperref{#1}{Assumption~\ref*{#1}}}
\newrefformat{ex}{\savehyperref{#1}{Example~\ref*{#1}}}
\newrefformat{fig}{\savehyperref{#1}{Figure~\ref*{#1}}}
\newrefformat{alg}{\savehyperref{#1}{Algorithm~\ref*{#1}}}
\newrefformat{rem}{\savehyperref{#1}{Remark~\ref*{#1}}}
\newrefformat{conj}{\savehyperref{#1}{Conjecture~\ref*{#1}}}
\newrefformat{prop}{\savehyperref{#1}{Proposition~\ref*{#1}}}
\newrefformat{proto}{\savehyperref{#1}{Protocol~\ref*{#1}}}
\newrefformat{prob}{\savehyperref{#1}{Problem~\ref*{#1}}}
\newrefformat{obs}{\savehyperref{#1}{Observation~\ref*{#1}}}
\newrefformat{prot}{\savehyperref{#1}{Property~\ref*{#1}}}
\newrefformat{claim}{\savehyperref{#1}{Claim~\ref*{#1}}}
\newrefformat{que}{\savehyperref{#1}{Question~\ref*{#1}}}
\newrefformat{op}{\savehyperref{#1}{Open Problem~\ref*{#1}}}
\newrefformat{fn}{\savehyperref{#1}{Footnote~\ref*{#1}}}

\def \epsilon {\varepsilon}

\usepackage{url}
\usepackage{graphicx}
\usepackage{mathtools}
\usepackage{footnote}
\usepackage{float}
\usepackage{xspace}
\usepackage{multirow}
\usepackage{xcolor}
\usepackage{wrapfig}
\usepackage{framed}
\usepackage{bbm}
\usepackage{footnote}
\usepackage{nicefrac}
\usepackage{makecell}
\usepackage{bm}
\makesavenoteenv{tabular}
\makesavenoteenv{table}

\renewcommand{\tilde}{\widetilde}
\renewcommand{\hat}{\widehat}

\newcommand{\LW}{\text{LW}}

\newcommand{\val}{\text{Val}}

\newcommand{\eps}{\epsilon}

\newcommand{\Reg}{\text{\rm Reg}}

\newcommand{\one}{\boldsymbol{1}}

\newcommand{\E}{\field{E}}

\newcommand{\wh}{\widehat}

\newcommand{\rbr}[1]{\left(#1\right)}
\newcommand{\sbr}[1]{\left[#1\right]}
\newcommand{\cbr}[1]{\left\{#1\right\}}

\usepackage{lipsum,booktabs}
\usepackage{amsmath,mathrsfs,amsfonts,bm,enumitem}
\usepackage{rotating}
\usepackage{pdflscape}
\usepackage{hyperref,url}

\usepackage{appendix}
\usepackage{multirow,makecell}

\renewcommand{\tilde}{\widetilde}
\renewcommand{\hat}{\widehat}

\def \E {\mathbb{E}}

\usepackage{mathtools}

\usepackage{tikz}

\begin{document}

\maketitle

\begin{abstract}
We study a game between autobidding algorithms that compete in an online advertising platform. Each autobidder is tasked with maximizing its advertiser's total value over multiple rounds of a repeated auction, subject to budget and return-on-investment constraints. We propose a gradient-based learning algorithm that is guaranteed to satisfy all constraints and achieves vanishing individual regret. Our algorithm uses only bandit feedback and can be used with the first- or second-price auction, as well as with any ``intermediate" auction format. Our main result is that when these autobidders play against each other, the resulting expected liquid welfare over all rounds is at least half of the expected optimal liquid welfare achieved by any allocation.  This holds whether or not the bidding dynamics converges to an equilibrium. 

\vspace{4mm}

\vspace{2mm}
\xhdr{Acknowledgements.}
This research has been done while SP and MZ were graduate students at MIT and USC, resp., and (partially) while they were research interns at Microsoft Research. The authors are grateful to Bach Ha (Microsoft Bing Ads) for many conversations that informed our perspective, and to Sidharth Satya for providing research support.

\vspace{2mm}
\xhdr{Version history.} January 2023: first version, with all theoretical results.\\ June 2024: added numerical experiments, revised presentation.\\
November 2024: this version, revised presentation.

\vspace{2mm}
\xhdr{Conference Publication.}
Presented at the 37th Conference on Learning Theory (COLT), 2024.\\
(Accepted as a full paper, published as an extended abstract to accommodate a journal submission.)

\end{abstract}


\newpage
\addtocontents{toc}{\protect\setcounter{tocdepth}{2}}
\tableofcontents
\newpage

\section{Introduction}




As the rules and algorithms governing online markets increase in complexity and scale, platforms are increasingly providing ML-powered interfaces to help users interact and navigate efficiently.  A prominent example is the rise of \emph{autobidding},
a service provided by advertising platforms to help advertisers automate their campaigns.
The advertiser only needs to specify high-level objectives and constraints.  A typical example might be ``maximize the number of clicks received, subject to spending at most \$1000 per day, at most \$2 per click on average, and no more than \$10 for any one click.''  This example encodes three different constraints on the outcome: a budget constraint,  an average return-on-investment (ROI) constraint,%
\footnote{A common alternative term is \emph{ROAS}, Return on Ad Spend.}
 and a maximum bid.
The autobidder then uses an online learning algorithm to tune a detailed advertising campaign so as to solve this optimization problem on the advertiser's behalf.
While each ad impression is sold by an auction, individual bids are managed by the autobidder.



Most major online ad platforms now feature integrated autobidding tools.
This popularity is owed in part to the effectiveness of online learning methods for bid tuning, which has recently received substantial attention in the academic literature.  Initial work focused on tuning bids subject to an aggregate budget constraint, a.k.a., budget pacing~\citep{Borgs-www07,balseiro2019learning,Conitzer-wine18,Conitzer-ec19}.
More recent work concerns ROI constraints
\citep{feng2022online,roi-balseiro2021robust,golrezaei2021bidding,roi-golrezaei2021auction,roi-li2022auto,roi-mehta2023auctions}. Thus, a variety of well-understood learning algorithms can be used by an autobidder to achieve vanishing regret in any stationary (or near-stationary) auction environment.  Correspondingly, autobidder interfaces supporting both budget and ROI constraints are now ubiquitous.


Since autobidders are now predominantly competing against each other, one must worry what happens
when {they}
interact.
What are the implications
for the overall health of the market as a whole,
in terms of aggregate objectives such as efficiency and {convergence}?
In particular, one needs to account for unintended emergent behaviors
that may arise when autobidders compete.
One solution would be to design autobidding algorithms that
always converge to equilibria of the ``bidding game" that they are playing against each other.
This would also address efficiency, if the corresponding equilibria are sufficiently efficient. The state of the art guarantees in this respect concern \emph{liquid welfare} (a standard notion of welfare under constraints, as discussed below), and \emph{pacing equilibrium},
a pure Nash equilibrium of the appropriately defined single-shot bidding game with budget and/or ROI constraints
\citep{Conitzer-wine18,Conitzer-ec19}. Specifically, for truthful auctions any pacing equilibrium attains expected liquid welfare
at least half of the optimum, and this bound is tight~\citep{Gagan-wine19,Babaioff-itcs21}. Unfortunately,
finding an equilibrium of this bidding game is PPAD-hard, even for the special case of {budget-constrained}
second-price auctions~\citep{Chen-ec21}.  We therefore should not expect to design an algorithm that is always guaranteed to jointly converge to an equilibrium when deployed by autobidders.

This leaves us with the challenge of analyzing the joint learning dynamics of the autobidders, without relying on convergence.
While this challenge may appear daunting, recent developments suggest cautious optimism.  For budget-constrained advertisers,~\citet{Gaitonde-itcs23} showed that the autobidding algorithm of~\citet{balseiro2019learning} does indeed generate high expected liquid welfare
even when the learning dynamics does not converge.  Their analysis is specific to budget constraints, and
does not appear to apply to any existing multi-constraint autobidding algorithm.
Nevertheless, it is natural to ask whether such results can be extended to the common scenario with both budget and ROI constraints, perhaps via new algorithms.




\xhdr{Our contributions.}
We present a novel algorithm for autobidding with budget and ROI constraints. While this by itself is not new -- prior work achieves vanishing regret under such constraints -- the critical new feature of our algorithm is an aggregate guarantee:
when multiple autobidders all deploy our algorithm, the resulting expected liquid welfare is at least half of the optimal. This matches the best possible bound for pacing equilibria.%
\footnote{\label{fn:pacing}
The following lower bound holds for any $\lambda>\tfrac12$: the liquid welfare of a pacing equilibrium cannot exceed $\lambda$ fraction of the optimum for all budget-constrained bidding games and all equilibria. This holds for second-price auctions as per ~\citet{Gagan-wine19,Babaioff-itcs21}, and also for first-price auctions due to a simple example, see \pref{app:pacing}. A matching upper bound holds for all truthful auctions and all pacing equilibria.}
However, our aggregate guarantee \emph{does not} rely on convergence to equilibrium; rather, like \citet{Gaitonde-itcs23} for budget-constrained bidders, we directly analyze the learning dynamics.%
\footnote{In particular, the connection to pacing equilibria is only informal. Equilibrium analysis is a common first step towards analyzing the dynamics, and equilibria results such as the lower bound in Footnote~\ref{fn:pacing} are commonly interpreted as informal benchmarks for the respective dynamics. However, such lower bounds apply formally only if the dynamics can converge to a worst-case pacing equilibrium (which is not necessarily true for all learning dynamics).}


We measure market efficiency via \emph{liquid welfare},
the maximum amount the agents are willing to pay for the allocations that they receive.  This is the appropriate notion of welfare in settings like ours where each agent's (i.e., autobidder's) goal is to
maximize value subject to monetary constraints.%
\footnote{When/if the advertisers' objective \emph{is} expressible in dollars, utilitarian welfare could also be a reasonable objective. However, strong impossibility results \citep{Shahar-icalp14} make it less suitable for theoretical study.}
Liquid welfare generalizes the standard notion of welfare for agents with quasi-linear preferences in money (or, equivalently, only a constraint on the maximum willingness-to-pay in each round). {It has been introduced for the special case of
budget constraints in \citet{Shahar-icalp14}, and has become a standard objective in the welfare analysis of constrained auctions since then
 \citep{Azar-wine17,Gagan-wine19,Babaioff-itcs21,Gaitonde-itcs23}.}



We also guarantee good performance for each individual autobidder.
(Such guarantees are crucial even when the platform's objective is overall market efficiency,
since otherwise advertisers might prefer to forego autobidding and place bids themselves.)
Specifically, we obtain vanishing regret in an adversarial environment, without any assumptions on the other agents. Our result holds relative to
the \emph{constraint-pacing sequence}: informally, a sequence of bids that maximize
expected value in each round under the time-averaged expected  constraints.
Note that vanishing-regret results are impossible against the standard benchmark of best fixed bid, even for the budget constraint only
\citep[][also see \pref{sec:related}]{balseiro2019learning}. The primary reason for that is the \emph{spend-or-save dilemma}: whether to spend the budget now or save it for later. A natural way to side-step this dilemma is to satisfy the expected constraints in each round, as in our benchmark.
Specialized to a stationary environment (where the other agents' bids are drawn from a fixed joint distribution), the benchmarks become equivalent and hence we obtain vanishing regret against the best fixed bid.


Our results hold for a broad class of auction formats, including first-price and second-price auctions, and allow impression {types} (which determine, e.g., click rates) to be drawn randomly in each round and potentially be correlated across agents.
Our algorithm
is guaranteed to satisfy all constraints ex post with probability $1$ (not just with high probability or with small expected violation). Further, it only requires bandit feedback from the underlying auction (i.e., only the outcome for the actual bid submitted, not the counterfactual outcomes for the  alternative bids).  This
is important
because even when advertising platforms provide sufficient feedback to infer the counterfactual slate of ad impressions (which they do not always), it can be difficult
to accurately model which ads in the slate
would be clicked by a user.


\xhdr{Our techniques.}
Our algorithm uses a non-standard variation of stochastic gradient descent (SGD).%
\footnote{SGD is a classic algorithm in online convex optimization \citep{Hazan-OCO-book}.}
When there is only a budget constraint, a common idea is to use SGD-based updates to learn the constraint-pacing bid: a bid that
exactly spends the budget in expectation.
With both budget and ROI constraints, one could strive to learn constraint-pacing bids for each constraint, and then aggregate these into a bidding strategy.
For example, \citet{Balseiro-BestOfMany-Opre} employ a primal-dual framework that
interpolates and places more weight on constraints that bind more tightly; \citet{feng2022online} apply this directly to the setting of ROI and budget constraints.

Our approach has a similar flavor, but aggregates the two per-constraint bids differently.
Each round, our autobidder myopically uses the smaller of the two constraint-pacing bids, then applies an SGD step to \emph{both} bids using the observed outcome.
It may seem counter-intuitive to update the larger bid using a gradient evaluated at the smaller bid, and indeed this breaks the usual convergence guarantees of SGD.
However, this approach maintains an important invariant: the multiplier for each constraint encodes the total slack (or violation) of that constraint up to the current round.
This invariant
lets us track outcomes while being agnostic to the details of the (potentially chaotic and non-convergent) bid dynamics.  It also {implies}
that all constraints are satisfied with certainty. 

When there is only a budget constraint, our algorithm specializes to the autobidding algorithm from  \citep{balseiro2019learning}, and our guarantees specialize to the regret and liquid welfare guarantees from \citet{Gaitonde-itcs23}.  Like that work, our aggregate guarantees employ an ``approximate First Welfare Theorem'' approach that charges any welfare loss from suboptimal allocations to revenue collected by the platform.  This is a common technique for equilibrium analysis, including pacing equilibrium, but extending this approach to non-convergent learning dynamics requires new ideas; see Appendix~\ref{app:pacing}.
While our approach to bounding liquid welfare
shares a common high-level strategy with~\citet{Gaitonde-itcs23},
handling the ROI constraint, and particularly both constraints jointly, introduces a variety of new technical challenges (discussed in \pref{sec: multi-cons}) and motivates our proposed algorithm.
We provide a detailed technical comparison in \pref{app:comparison}.

\xhdr{Discussion.}
We posit that all advertisers run the same fixed algorithm, a standard assumption in theoretical results on multi-agent learning. One motivation is that the algorithm is chosen and implemented by the platform, rather than directly controlled by the advertisers.  Beyond  vanishing regret, we do not explicitly consider advertisers' incentives to use this algorithm. However, the advertisers are not likely to switch to a DIY solution without a substantial advantage: indeed, the platform-provided autobidder does not require implementation efforts and may have better access to the platform's internal statistics and estimates.

The bids placed by an autobidder in our model scale linearly with realized impression value,
following most prior work on online bidding.%
\footnote{Specifically, all prior work on online bidding that we are aware of, except \citet{StandardAuctions-ec22}.}
The linear scaling is inevitable if the impression values are not observable in advance (as in our model). 
It is optimal for bidders for truthful auctions under budget constraints \citep{balseiro2019learning,Babaioff-itcs21}, but not more generally. Accordingly, our individual guarantees for the general case only compare against linear
bids.


\xhdr{Numerical experiments.}
We provide numerical experiments to complement our theoretical results. We focus on a standard ``multi-player" environment when all advertisers run the same algorithm. Our liquid welfare guarantees apply to this environment, but empirical performance could potentially be even better. Moreover, while our individual regret bounds apply to a general adversarial environment, they do \emph{not} meaningfully specialize to the multi-player environment (see \pref{rem:multiplayer-regret} in \pref{sec: multi-cons-indi}). So, individual regret for this environment is only studied via experiments.



We consider several simulated problem instances with budget and ROI constraints. We track several metrics: constraint slackness/violation, liquid welfare, and individual regret, as well as the dynamics of the multipliers. To get a sense of the problem difficulty, we also consider other bidding algorithms (in each experiment, all bidders are assigned the same algorithm). Specifically, we consider BOMW, a ``smart" algorithm from  \citep{Balseiro-BestOfMany-Opre,feng2022online}, and two ``naive" baselines: the greedy algorithm and the epsilon-greedy algorithm.%
\footnote{Both are well-known ``templates" in multi-armed bandits. In particular, the greedy algorithm ignores the need for exploration, and the epsilon-greedy algorithm (in our setting) ignores the non-stationarity and the constraints.}

Our high-level findings are as follows.
First, our algorithm satisfies all constraints and achieves high liquid welfare relative to the baselines,
whereas the ``naive" baselines fail due to large constraint violations.
%
Second, the dynamics does not always converge to a stationary state
(within the timeframe considered)
for our algorithm nor for BOMW. This further motivates the study of aggregate guarantees such as ours that bypass convergence, and individual guarantees such as ours that go beyond the stationary environment. We observe strong algorithm performance regardless of whether the dynamics has converged.
Third, our algorithm appears to have vanishing individual regret relative to the best-in-hindsight bid, despite non-convergence of the dynamics.  We empirically estimate a best-fit parameter $\RegPow>0$ such that observed regret evolves according to $T^{\RegPow}$, and we obtain an empirical estimate bounded away from $1$ in all instances simulated.

\section{Further Discussion of Related Work}
\label{sec:related}

\noindent\textbf{Online bidding.}
Our work builds on the recent literature analyzing online algorithms
for bidding under constraints. For the special case of a budget constraint, \citet{Borgs-www07} analyze bidding dynamics under a multiplicative update rule and establish convergence for first-price auctions. \citet{balseiro2019learning} consider a different update rule under second-price auctions and show that it converges under some additional convexity assumptions and guarantees vanishing individual bidder regret. \citet{Balseiro-BestOfMany-Opre} consider a variation of this approach using online mirror descent (OMD) and extend the individual guarantees to repeated truthful auctions.
Notably, their regret bound applies to adversarial environments, with a loss that grows with the deviation from the stationary environment.


{For budget and ROI constraints, \citet{gao2022bidding} propose and evaluate a dual-based optimization framework, without any provable guarantees. \citet{golrezaei2021bidding} achieve low regret in a stationary environment with bounds on expected constraint violations. \citet{feng2022online} extend the OMD-based approach from \citet{Balseiro-BestOfMany-Opre} to achieve vanishing regret while satisfying constraints with probability 1. We emphasize that these papers do not provide any aggregate guarantees, or any individual guarantees beyond the stationary environment. \citet{golrezaei2021bidding,feng2022online} focus on, resp., repeated second-price auctions and repeated truthful auctions. They do, however, achieve better regret rates in the stationary environment: $T^{1/2}$ regret in \citet{golrezaei2021bidding,feng2022online} vs. $T^{7/8}$ regret in our paper.}



\xhdr{Equilibrium analysis.}
A line of work studies \emph{pacing equilibria}: market equilibria in a single-shot game that abstracts repeated auctions when bidders have global constraints such as budgets and/or ROI. \citet{Conitzer-wine18} introduced such equilibria and characterized them for second-price auctions with budget constraints. For second-price auctions, the expected liquid welfare obtained at any pacing equilibrium is at least half of the optimal liquid welfare; this holds for a broad class of constraints including budget and ROI constraints \citep{Gagan-wine19}. \citet{Babaioff-itcs21} provided a similar 2-approximation result in settings with ``soft constraints'' where agents have a separable and convex disutility for spending money. However, finding a pacing equilibrium under budget constraints is PPAD-hard~\citep{Chen-ec21}. Recall that our analysis of liquid welfare does not rely on the equilibrium analysis, since it does not rely on convergence to equilibrium.

\citet{Conitzer-ec19} extended pacing equilibria to first-price auctions, showed that the equilibrium is essentially unique, and analyzed its properties. \citet{StandardAuctions-ec22} considered an alternative equilibrium notion for a broad class of auctions,
in which agents are not constrained to pacing but instead can make their bids arbitrarily contingent on realized impression values.  They present a revenue-equivalence result to bound liquid welfare at any equilibrium subject to a budget constraint.




\xhdr{Learning theory.}
Repeated bidding under budget is a special case of \emph{bandits with knapsacks} (BwK), a multi-armed bandit problem under global constraints on resource consumption
\citep{BwK-focs13,AgrawalDevanur-ec14-OpRe,AdvBwK-focs19}, see Chapter 10 in \citep{slivkins-MABbook} for a survey.
BwK problems in adversarial environments do not admit
vanishing regret bounds against standard benchmarks: instead,
one is doomed to approximation ratios,
even in relatively simple examples \citep{AdvBwK-focs19}. A similar impossibility result is derived in
\citep{balseiro2019learning} specifically for repeated budget-constrained bidding in second-price auctions. These results hinge on the \emph{spend-or-save dilemma}: an algorithm does not know in advance whether to spend the budget now or save it for the future, and inevitably makes a wrong choice for some variant of the future. Prior work on BwK
is not concerned with aggregate guarantees such as ours.


Convergence of learning algorithms in repeated games is extensively studied, yet not well-understood.
When algorithms have vanishing regret in terms of cumulative payoffs, the \emph{average play} (time-averaged distribution over chosen actions) converges to a (coarse) correlated equilibrium~\citep{Aumann-74,Moulin-78,HartMasCollel-econometrica00}, and this implies welfare bounds for various auction formats in the absence of budget or ROI constraints~\citep{DBLP:journals/jair/RoughgardenST17}.  In contrast, for repeated auctions with budgets, low individual regret on its own does not imply any bounded approximation for liquid welfare \citep{Gaitonde-itcs23}.
Convergence in the last iterate is even more challenging. While strong negative results are known even for two-player zero-sum games \citep{Piliouras-ec18,Piliouras-soda18,Piliouras-colt19}, a recent line of work \citep{Daskalakis-iclr18,Daskalakis-itcs19-lastIterate,Daskalakis-neurips20-lastIterate,Haipeng-iclr21-lastIterate} achieves last-iterate convergence under full feedback and substantial convexity-like assumptions, using two specific regret-minimizing algorithms. To the best of our understanding, these positive results do not apply to repeated auctions with budget or ROI constraints.

\section{Model and Preliminaries}\label{sec: model}

We study a repeated auction game played by a collection of $n$ bidding agents (i.e., autobidders).  At each round $t = 1, 2, \dotsc, T$, the seller (or platform) has a single unit of good available to sell.  An \textit{allocation profile} is a vector $\boldsymbol{x} = (x_{1}, \dotsc, x_{n}) \in X \subseteq [0,1]^n$ where $x_{k}$ is the quantity of the good allocated to agent $k$ and $X$ is any downward-closed set of feasible allocations.%
\footnote{Meaning: $x\in X \Rightarrow x'\in X$ for any two vectors $x,x'\in [0,1]^n$ with $x'_i\leq x_i$ $\forall\,i\in[n]$.
The canonical case is integer allocation, when $X$ consists of all unit vectors, but
$X$ can also allow fractional allocations.}
An \emph{allocation sequence} is a sequence of allocation profiles $(\boldsymbol{x}_{1},\ldots,\boldsymbol{x}_{T})$ where $\boldsymbol{x_{t}} = (x_{1,t}, \dotsc, x_{n,t})$ is the allocation profile at round $t$. (The notation is summarized in \pref{app: notation}.)

The good available for sale at a given round $t$ has a click probability $\quality_{k,t} \in [0,1]$ for each agent $k$.\footnote{We could alternatively think of $\quality_{k,t}$ as a conversion rate, an expected revenue lift, or any other driver of realized impression value.  For clarity of exposition we will use the language of clicks from now on.}
The tuple $\mathbf{\quality}_t = (\quality_{1,t}, \dotsc, \quality_{n,t})$ is drawn from a fixed distribution $\distrC$ independently across time periods.  Note that $\distrC$ {can be arbitrarily correlated across agents}.



\xhdr{Auctions.}
At each round $t$, the good is allocated via an auction that proceeds as follows.  Each agent $k$ submits a bid $\beta_{k,t} \geq 0$, which can be interpreted as a bid (in dollars) per click.
All agents submit bids simultaneously.  The $\mathbf{\quality}_t$ tuple is then realized, which determines each agent's effective bid $b_{k,t} = \quality_{k,t} \beta_{k,t}$.\footnote{Equivalently, the bidder observes $\quality_{k,t}$, but is restricted to choosing an effective bid $b_{k,t}$ that is linear in $\quality_{k,t}$.}
The auction is defined by an allocation rule $\boldsymbol{x}$ and a payment rule $\boldsymbol{p}$, where $\boldsymbol{x}(\boldsymbol{b}) \in [0,1]^n$ is the allocation profile generated under a bid profile $\boldsymbol{b}$, and $p_k(\boldsymbol{b}) \geq 0$ is the payment made by agent $k$.  Allocation and payment rules are always weakly monotone in bids, meaning that $x_k(b_k, \mathbf{b}_{-k})$ and $p_k(b_k, \mathbf{b}_{-k})$ are weakly increasing in $b_k$ for any $\mathbf{b}_{-k}$.  Given an implied realization of bids, we will often write $x_{k,t}$ and $p_{k,t}$ to denote the allocation and payment of agent $k$ in round $t$.

Auction rules satisfy the following two properties. Winners (i.e., agents with non-zero allocation) are always selected from among agents with the highest effective bid.
Also, each agent's payment per unit received, $p_{k,t}/x_{k,t}$, lies between the highest and second-highest effective bids. This includes the first-price and second-price auctions, as well as anything ``in between."
\footnote{Some of our results actually apply more generally to \emph{core auctions}; we define these in~\pref{sec:liquid.welfare}.}



\xhdr{Objective.}
Each agent $k$ 
maximizes $\sum_k \quality_{k,t} x_{k,t}$  (i.e., expected clicks)
subject to the following: 

\begin{description}
\setlength{\itemsep}{0pt}
\setlength{\parsep}{0pt}
\setlength{\topsep }{0pt}
\item[Bid constraint ($\theta_k$).]
In each round $t$,
the bid cannot exceed $\theta_k$:
        $\beta_{k,t} \leq \theta_k$.
\item[ROI constraint ($w_k$).]
The total payment cannot exceed $w_k$ per click:
    $\sum_t p_{k,t} \leq w_k \sum_t \quality_{k,t} x_{k,t}$.
\item[Budget constraint ($B_k$).]
The total payment
 cannot exceed $B_k$: $\sum_t p_{k,t} \leq B_k$.
\end{description}




All constraints bind ex post, and must be satisfied on every realization.
%
Note there always exists an agent strategy that guarantees all constraints will be satisfied.
The first constraint is always satisfied if $\beta_{k,t} \leq \theta_k$ for all $t$;
the second constraint is always satisfied if $\beta_{k,t} \leq w_k$ for all $t$;
and the third constraint is always satisfied if $\beta_{k,t} \leq B_k / T$ for all $t$.  Bidding the minimum of these
would therefore necessarily satisfy all constraints, though of course this may result in low objective value.

We write $\rho_k = B_k/T$ for agent $k$'s per-round budget constraint. For convenience, define $v_{k,t} = \theta_k c_{k,t}$ for the maximum allowed effective bid for agent $k$ in round $t$.  Since the ROI constraint would be implied by the bid constraint if $w_k > \theta_k$, we will assume without loss of generality that $w_k \leq \theta_k$, and define $\gamma_k = \theta_k/w_k \geq 1$.  The ROI constraint can then be rewritten as
    $\gamma_k \sum_t p_{k,t} \leq \sum_t v_{k,t} x_{k,t}$.
Finally, following the literature, we define multiplier $\mu_{k,t} = (\theta_k/\beta_{k,t})-1$ so that $\beta_{k,t} = \theta_k/(\mu_{k,t}+1)$.  We then think of agent $k$'s problem as choosing multiplier $\mu_{k,t} \geq 0$, where $0$ corresponds to the maximum allowable bid and larger values of $\mu_{k,t}$ correspond to smaller bids.

\begin{remark}
The ROI constraint is trivially satisfied when $\gamma_k = 1$. (This is because
    $p_{k,t}/x_{k,t} \leq c_{k,t}\,\beta_{k,t} \leq c_{k,t}\, \theta_k$.)
Put differently, $\gamma_k = 1$ corresponds to the budget-only case.
\end{remark}

Each autobidding agent $k$ can be equivalently thought of as maximizing $\sum_t v_{k,t}\cdot x_{k,t}$ subject to the constraints. This motivates us to refer to each $v_{k,t}$ as a \emph{value} for agent $k$ at round $t$.
However, we emphasize that these ``values" are independent of the advertiser's {actual} utility model, to which we are agnostic.  Rather, they {merely parameterize the autobidders' objective}.
{As a shorthand, we also define \emph{value profiles}
    $\mathbf{v}_t := (v_{1,t}, \dotsc, v_{n,t})$
for each round $t$, which are drawn independently from some fixed distribution $F$ (induced by $\distrC$). The \emph{value sequence} is the sequence
    $\mathbf{v} = \rbr{\mathbf{v}_1,\ldots,\mathbf{v}_t}$.

\xhdr{Liquid welfare} is a welfare notion for agents with payment constraints which measures agents' \emph{maximum willingness to pay} for the received allocation. For our setting, it is defined as follows:


\begin{definition}
\label{def:liquid_welfare}
Fix a value sequence $\mathbf{v} = \rbr{\mathbf{v}_1,\ldots,\mathbf{v}_T} = (v_{k,t}) \in [0,1]^{nT}$ and a  feasible allocation sequence $\mathbf{x}= \rbr{\mathbf{x}_1,\ldots,\mathbf{x}_T}= (x_{k,t}) \in X^T$. The corresponding \emph{liquid value} obtained by agent $k$ is
     $W_{k}(\mathbf{x},\mathbf{v})
     =\min\left\{B_k,\, \frac{1}{\gamma_k} \sum_{t=1}^T v_{k,t}x_{k,t}\right\}$,
 and \emph{liquid welfare} is $W(\mathbf{x},\mathbf{v}):=\sum_{k=1}^n W_{k}(\mathbf{x},\mathbf{v}).$
 \end{definition}


We emphasize that liquid welfare depends on the allocations, but not on the agents' payments.  {However, it follows immediately that an agent's total payment cannot exceed her liquid value.}


\begin{observation}
\label{obs:LW.revenue}
Fix any sequence of value profiles and outcomes (allocations and payments) such that the bid, budget, and ROI constraints are satisfied for a given agent $k$. {In the notation from  \pref{def:liquid_welfare}, the total payment of agent $k$ cannot exceed $W_k(\mathbf{x},\mathbf{v})$.}
\end{observation}

Our objective of interest is the \emph{expected} liquid welfare, over any randomness in the value sequence and the agents' autobidding algorithms.
Since the bids in one round can depend on allocations from the previous rounds, we define
a mapping $\Phi$ from the entire value sequence $\mathbf{v}$ to an allocation sequence
    $\mathbf{x} = \Phi(\mathbf{v})$;
call $\Phi$ an \emph{allocation-sequence rule}. (Note that it abstracts both the autobidding algorithm and the underlying auction. Treat $\Phi$ as randomized if either the algorithm or the auction are.) Then the expected liquid welfare is
    $W(\Phi,F) := \E_{\mathbf{v}}\sbr{W(\Phi(\mathbf{v}),\mathbf{v})}$,
where the value sequence $\mathbf{v}$ drawn according to distribution $F$.


\section{Warmup: ROI Constraints and Unlimited Budget}
\label{sec:just_ROI}




Consider unlimited budgets, i.e., $B_k = \infty$ for all $k$. While bidding algorithms are already known for this problem,
we present a new algorithm and emphasize some properties 
that will be crucial for the general setting we consider in~\pref{sec: multi-cons}.
%
%
%
%
%
%
As we focus on a single agent, we will drop the subscripts $k$ from our notation for the remainder of this section.
In the absence of budget constraints, the agent's goal is to maximize $\sum_t v_t x_t$ subject to the ROI constraint $\sum_t v_t x_t - \gamma \sum_t p_t \geq 0$.
\begin{example}\label{ex:warmup}
Consider a repeated second-price auction with $v_t = 1$ for every $t$ and $\gamma = 2$.  Then the agent's ROI constraint is $\sum_t x_t - 2 \sum_t p_t \geq 0$. Suppose the competing bids are stochastic and stationary, with the highest competing bid $\bmax$ either $\nicefrac{1}{4}$ or $\nicefrac{3}{4}$ each round with equal probability.

Suppose that the bidding agent chooses $\mu_t = 0$ for every round $t$, so it bids $b_t = v_t = 1$.  Then the agent wins every round and pays either $\nicefrac{1}{4}$ or $\nicefrac{3}{4}$, for an expected payment of $\nicefrac{1}{2}$.  The constraint is satisfied in expectation but may be violated on some realizations.  Indeed, in the (unlikely) event that $\bmax=\nicefrac{3}{4}$ every round, the only way to satisfy the ROI constraint would be to lose every round. However, if $\bmax=\nicefrac{1}{4}$ in at least half of the rounds then it is optimal to win every round.
\end{example}

%
%
%

Setting $\mu_t = \gamma-1$ in every round (i.e., bidding $v_t/\gamma$) is guaranteed to satisfy the ROI constraint (see~\pref{sec: model}).
However, the optimal choice of $\mu_t$ may be lower, as in \pref{ex:warmup}.  
Next, we show that ROI is monotone in $\mu_t$ for our auctions: higher bids result in lower average ROI.
This lets us think of the agent as trying to reduce $\mu$ as much as possible subject to the ROI constraint.

\begin{lemma}
\label{lem: warmup_1}
Fix any auction in our class, any agent $k$, and any bids $\textbf{b}_{-k}$ of the other agents, and write $x_t(\mu)$ and $p_t(\mu)$ for the allocation and payment that results when agent $k$ selects multiplier $\mu$.  Then $v_t x_t(\mu) - \gamma p_t(\mu)$ is weakly increasing in $\mu$, for $\mu \in [0, \gamma - 1]$.
\end{lemma}

This motivates us to consider
\pref{alg:bg_roi}, which initially takes the safe action $\mu_1 = \gamma - 1$ but updates $\mu_t$ online in response to auction feedback.
In each round $t$ it places bid $b_t = v_t / (1 + \mu_t)$.\footnote{With one small caveat: $\mu_t$ could be negative in some rounds, in which case we treat it as $0$ when setting the bid $b_t$.}
If the observed allocation $x_t$ and payment $p_t$ in round $t$ are such that $v_t x_t > \gamma p_t$ then the ROI constraint is  satisfied with room to spare.  This suggests the bid was lower than necessary, so the algorithm reduces $\mu$ by an amount proportional to $v_t x_t - \gamma p_t$.  Likewise, if $v_t x_t < \gamma p_t$, then the ROI constraint was violated in round $t$ so the algorithm responds by increasing $\mu$ proportionally to the violation.  This can be interpreted as stochastic gradient descent (SGD): if $\mu_t$ is such that the ROI constraint is satisfied in expectation then the expected update is $0$.  We will make this connection with SGD more precise in~\pref{sec: multi-cons-indi}.

\begin{algorithm2e}[h]
\SetAlgoNoEnd
\caption{Bidding Under ROI Constraint}\label{alg:bg_roi}
\vspace{2mm}

Input: ROI constraint parameter $\gamma$.\\
Initialization: $\mu_1 = \gamma - 1$ and learning rate $\eta>0$.\\
\For{$t = 1, 2, \cdots , T$}{
    Set bid $b_t = \frac{v_t}{1 + \max\{\mu_t, 0\}}$

    Observe allocation $x_t\in [0,1]$ and payment $p_t$

    Update the multiplier $\mu_{t+1} = \mu_t - \eta(v_t x_t - \gamma p_{t} )$
}
\end{algorithm2e}

\pref{alg:bg_roi} appears myopic at first glance, always updating its bids in response to the latest outcome.
However, we note that $\mu_t$ implicitly encodes the status of the aggregate ROI constraint up to round $t$.  Indeed,
an immediate implication of the update rule is that $\mu_t$ is $\mu_1$ minus a term proportional to $\sum_{\tau < t}(v_\tau x_\tau - \gamma p_\tau)$, which is precisely the aggregate slack (or violation) of the ROI constraint up to time $t$.  A small value of $\mu_t$ (i.e., a high bid) therefore occurs only if there is substantial slack in the ROI constraint up to round $t$.

What does this mean for the performance of \pref{alg:bg_roi}?  Note that the bidding agent may receive less than the desired return on investment $\gamma$ in any given round.  However, as we now show, the algorithm is guaranteed, with probability $1$, to satisfy the ROI constraint in aggregate over its $T$ rounds (and indeed, over any prefix of the $T$ rounds).  Intuitively,
as the ROI constraint gets closer to being violated in aggregate, $\mu_t$ gets closer to $\mu_1 = \gamma-1$,  the ``safe'' choice at which the ROI constraint will be satisfied each round.
\begin{lemma}
\label{lemma:ROI.constraint.ex.post}
Fix any (possibly adversarial) mapping from sequences of bids to sequences of allocations and payments such that $p_t \leq b_t x_t$ for every $t$, and suppose $\eta \leq 1/\bar{v}$.  Then for the allocations and payments resulting from applying \pref{alg:bg_roi}, we have $\sum_t v_t x_t \geq \gamma \sum_t p_t$.
\end{lemma}

An immediate corollary of~\pref{lemma:ROI.constraint.ex.post} is that, for every input sequence, the sequence of multipliers $\{\mu_t\}$ will satisfy $\mu_{t} \leq \gamma-1$ for every $t$.\footnote{$\mu_1 - \mu_t$ is proportional to the slack in the ROI constraint up to time $t$, which by \pref{lemma:ROI.constraint.ex.post} is never negative.}
As we show in the next section, this will imply high liquid welfare when multiple agents use (a generalization of) this algorithm.
%
%
%
As it turns out,~\pref{lemma:ROI.constraint.ex.post} can also be used to show that~\pref{alg:bg_roi} achieves vanishing regret relative to the best choice of $\mu$ in hindsight.  We prove this formally in the next section in a more general setting.
For now let us briefly describe the intuition.  Since the ROI constraint is satisfied with probability $1$ by \pref{lemma:ROI.constraint.ex.post}, any loss in value must come from bidding too low relative to the optimal fixed strategy in hindsight, say $\mu^*$.  However, the expected update to $\mu_t$ turns out to be the gradient of a function that, on the range $[\mu^*, \gamma-1]$, is (a) convex
and (b) closely related to the value function.  Standard SGD analysis then 
bounds the total loss due to rounds where $\mu_t$ is larger than $\mu^*$. 

\section{Bidding under ROI and Budget Constraints}\label{sec: multi-cons}

We now turn to the general problem with both ROI and budget constraints. We study an extension of our previous algorithm to this setting, listed below as \pref{alg:bg_roi_bud}. The algorithm now keeps track of two multipliers, $\mu^R$ and $\mu^B$, corresponding to the ROI and budget constraints respectively.  At each round, the algorithm will place a bid using whichever of the multipliers is more conservative; i.e., whichever results in the lower bid.  Each of the multipliers is then updated {according to the realized allocation and payment} \emph{as if that multiplier was used to set the bid} (even if it wasn't).
Multiplier $\mu^R_{k,t}$ is updated in the same way as ~\pref{alg:bg_roi}.  The idea behind our update rule for multiplier $\mu^B_{k,t}$ is similar: each round we compare the observed payment $p_{k,t}$ with $\rho_k = B_k/T$, the target per-round payment according to budget constraint $B_k$, and we update $\mu^B_{k,t}$ proportional to the difference where $\eta_{k,R},\eta_{k,B}>0$ is the resp. learning rate:
\begin{align*}
    \mu^R_{k,t+1} = \mu^R_{k,t} - \eta_{k,R}(v_{k,t} x_{k,t} - \gamma_k p_{k,t}), \nonumber \qquad \mu^B_{k,t+1} = \mu^B_{k,t} - \eta_{k,B}(\rho_k - p_{k,t}).
\end{align*}

%
%


\begin{algorithm2e}[t]
\SetAlgoNoEnd
\caption{Bidding Under ROI and Budget Constraints (for agent $k$)}\label{alg:bg_roi_bud}
\vspace{2mm}

Input: per-round budget constraint $\rho_k>0$ and ROI constraint parameter $\gamma_k$.

{Initialization: $\mu^R_{k,1} = \gamma_k - 1$, $\mu^B_{k,1} = \frac{\overlinev}{\rho}-1$ and learning rate $\eta_{k,R}, \eta_{k,B}>0$}.

\For{$t = 1, 2, \cdots , T$}{
    Calculate $\mu_{k,t} = \max \{\mu^R_{k,t}, \mu^B_{k,t}, 0\}$.

    Receive value $v_{k,t}$ and set the bid $b_{k, t} = \frac{v_{k,t}}{1 + \mu_{k,t}}$.

    Receive the allocation $x_{k,t}\in [0,1]$ and the payment $p_{k,t}$.

    Update the ROI-multiplier $\mu^R_{k,t+1} = \mu^R_{k,t} + \eta_{k,R}(\gamma_k p_{k,t} - v_{k,t}x_{k,t})$.

    Update the budget-multiplier $\mu^B_{k,t+1} = \mu^B_{k,t} + \eta_{k,B}( p_{k,t} - \rho_k)$.
}
\end{algorithm2e}


When the budget constraint is infinite (i.e., $B_k$ and hence $\rho_k$ is $+\infty$), this update rule yields $\mu_{k,t}^B < 0$ for every round $t$.  Thus, this algorithm reduces to ~\pref{alg:bg_roi}.  On the other hand, if $\gamma_k = 1$ (hence, no ROI constraint), then
$\mu_{k,t}^R \leq 0$ for every round $t$ (since $p_{k,t} \leq b_{k,t}x_{k,t} \leq v_{k,t}x_{k,t}$).  The resulting algorithm is nearly identical to the one in~\citet{balseiro2019learning} for bidding subject to a budget constraint.  We can therefore view ~\pref{alg:bg_roi_bud} as a generalization of both algorithms.


The key insight behind ~\pref{alg:bg_roi_bud} is that the multipliers $\mu_{k,t}^R$ and $\mu_{k,t}^B$ encode the cumulative slack in the ROI and budget constraints, respectively, up to time $t$.  
Similar to our analysis of \pref{alg:bg_roi}, it follows that ~\pref{alg:bg_roi_bud}
satisfies all of its constraints ex post with probability $1$:
\begin{lemma}
\label{lemma:ROI.budget.constraint.ex.post}
Fix any agent $k$ and any (possibly adversarial) mapping from sequences of bids to sequences of allocations and payments such that $p_{k,t} \leq b_{k,t} x_{k,t}$ for every round $t$. Assume $\eta_{k,R} \leq \frac{1}{\overlinev}, \eta_{k,B}\leq\min\{\frac{1}{\rho_k}, \frac{1}{\overlinev}\}$.  Then
\pref{alg:bg_roi_bud} satisfies
$\sum_t v_{k,t} x_{k,t} \geq \gamma_k \sum_t p_{k,t}$ and $\sum_t p_{k,t} \leq B_k$.
\end{lemma}

Also, similar to~\pref{alg:bg_roi}, another implication of this interpretation of $\mu_{k,t}^R$ and $\mu_{k,t}^B$ is that these multipliers are never higher than their ``safe'' levels. 


\begin{lemma}\label{lem:no more than gamma}
\pref{alg:bg_roi_bud} with learning rates $\eta_{k,R}\leq \frac{1}{\overlinev}$ and $\eta_{k,B}\leq \frac{1}{\overlinev} $ satisfies $\mu_{k,t}^{R}\leq \gamma_k -1$ and $\mu_{k,t}^B\leq \frac{\overlinev}{\rho_k}-1$
or all rounds $t\in[T]$.
\end{lemma}

The remainder of this section is dedicated to establishing our aggregate liquid welfare and individual regret guarantees for~\pref{alg:bg_roi_bud}.  In \pref{sec:liquid.welfare} we prove that in expectation over the realization of values,~\pref{alg:bg_roi_bud} always obtains at least half of the optimal liquid welfare in hindsight.  Then in \pref{sec: multi-cons-indi} we establish that the algorithm also satisfies strong individual regret guarantees, even in non-stochastic settings where the optimal bid sequence has bounded path length.



\subsection{Liquid Welfare Analysis}
\label{sec:liquid.welfare}

We prove that when all agents employ ~\pref{alg:bg_roi_bud} the resulting expected liquid welfare is approximately optimal.
We actually show that the expected liquid welfare is at least half of the optimal \emph{ex-ante} liquid welfare (defined in~\pref{def:liquid}), which is an agent's willingness to pay for her expected allocation sequence. This is a stronger benchmark compared to the optimal expected liquid welfare
due to Jensen's inequality (see~\pref{lem: ex_ante_motiv} in~\pref{app:aux}).





\begin{definition}\label{def:liquid}
For any distribution $F$ over valuation profiles and any allocation rule
    $\mathbf{y}: [0, \overline{v}]^{n} \rightarrow X$,
the \textit{ex-ante liquid welfare} is
    $\overline{W}(\mathbf{y}, F) := \sum_{k\in[n]} \overline{W}_k(\mathbf{y}, F)$,
where for each agent $k\in[n]$
    \begin{align}
        \overline{W}_k(\mathbf{y}, F) := T \times \min\left\{\rho_k,\; \tfrac{1}{\gamma_k}\; \mathbb{E}_{\mathbf{v} \sim F} [y_k(\mathbf{v})\, v_{k}] \right\}.
    \end{align}
\end{definition}


Recall that to this point we have focused our attention on single-item auctions.  Our liquid welfare bound will actually apply to the following more general class of \emph{core auctions}.


\begin{definition}\label{def:core}
Given any downward-closed set $X \subseteq [0,1]^n$ of feasible allocations, an auction with allocation rule $x \colon [0, \overline{v}]^n \to X$ and payment rule $p \colon [0, \overline{v}]^n \to \mathbbm{R}_{\geq 0}^n$ is a \emph{core auction} if it is
\begin{OneLiners}
\item Welfare-maximizing: $\mathbf{x}(\mathbf{v}) \in \arg\max_{\mathbf{x} \in X} \{\sum_i v_i(\mathbf{x})\}$
\item Individually rational: $p_i(\mathbf{v}) \leq v_i(\mathbf{x}(\mathbf{v}))$ for all $i$
\item Deviation-proof: for all $S \subseteq [n]$ and $\mathbf{y} \in X$, $\sum_{i \not\in S} p_i(\mathbf{v}) \geq \sum_{i \in S}(v_i(y_i) - v_i(x_i(\mathbf{v})))$
\end{OneLiners}
\end{definition}
Core auctions include first-price and second-price single-item auctions, but also more general formats like generalized second-price auctions for multiple slots and separable click rates (see~\cite{Gaitonde-itcs23} for further discussion).
%
%
We are ready to state the main result of this subsection, using the notation of allocation-sequence rule $\Phi$ and expected liquid welfare $W(\Phi,F)$ from \pref{sec: model}.

\begin{theorem}
\label{thm:liquid_welfare_roi_budget}
    {Fix any core auction as defined in \pref{def:core} and any distribution $F$ over agent value profiles. Suppose all agents bid by employing~\pref{alg:bg_roi_bud} with $\max\{\eta_{k,R},\eta_{k,B}\}\leq \frac{\overlinev}{\overlinev+\rho_k}\frac{\sqrt{\log(\overlinev n T)}}{\sqrt{T}}$. Write $\Phi$ for the corresponding allocation-sequence rule.
   Then for any allocation rule $\mathbf{y}: [0, \overline{v}]^{n} \rightarrow X$}, the expected liquid welfare $W(\Phi,F)$ satisfies
    \begin{align}
        {W(\Phi,F)}
        \geq \tfrac12\;\overline{W}(\mathbf{y}, F) - \mathcal{O}\rbr{n\overline{v} \sqrt{T \log (\overline{v} n T)}}.
    \end{align}
\end{theorem}

\xhdr{Proof Intuition.}%
\footnote{Since our approach shares common elements with a liquid welfare bound for budget constraints due to \citet{Gaitonde-itcs23}, we put particular emphasis on novel challenges and ideas; see \pref{app:comparison} for a more detailed comparison.}
Consider the liquid welfare (LW)  obtained by some agent $k$ over all $T$ rounds.  By definition, this LW is either the agent's budget $B_k$ or the sum of ROI-scaled gained values $\frac{1}{\gamma_k}\sum_t v_{k,t} x_{k,t}$.  The former case is easy: since $B_k$ is an upper bound on ex-ante LW, if agent $k$'s LW is $B_k$, then this is at least as good as the benchmark.  The difficulty lies in the latter case.

To bound $\frac{1}{\gamma_k}\sum_t v_{k,t} x_{k,t}$, we consider the progression of the bidding multiplier $\mu_{k,t}$ over rounds $t = 1, 2, \dotsc, T$.  The multiplier may drift up and down over time and may not converge.  We distinguish between rounds when $\mu_{k,t}$ lies above $\gamma_k - 1$ and rounds when $\mu_{k,t}$ lies below $\gamma_k - 1$.

If $\mu_{k,t} < \gamma_k - 1$
then agent $k$ is bidding $\geq v_{k,t} / \gamma_k$ on round $t$.
So even if agent $k$ loses in this round,
the winning bidder(s) must be paying $\geq v_{k,t} / \gamma_k$.  We can charge any loss in LW  against the total revenue collected, which (by~\pref{obs:LW.revenue}) is itself at most the LW.

On the other hand, in any round $t$ where $\mu_{k,t} > \gamma_k - 1$,
we know that $\mu_{k,t} = \mu_{k,t}^B$.  This is because  the ROI multiplier $\mu_{k,t}^R$ never lies above $\gamma_k - 1$ according to~\pref{lem:no more than gamma}.  Thus, over any contiguous interval of rounds in which $\mu_{k,t} > \gamma_k - 1$, it must be the budget multiplier that is determining the bid of agent $k$.  This allows us to employ an insight due to~\citet{Gaitonde-itcs23}: since any such interval must begin and end close to the threshold $\gamma_k - 1$, the update rule for $\mu_{k,t}^B$ implies that the total spend over the (say) $t$ rounds of that interval is very close to $t \times \rho_k$.  As the optimal ex-ante LW cannot be more than $\rho_k$ per round, the obtained LW must be comparable to the optimal LW over this interval.  Thus, in every case, we can relate the obtained LW to the benchmark.

There are some technical challenges to formalizing this intuition.  Our very first step was to condition on whether agent $k$'s total LW is determined by her budget or by her ROI-scaled gained value. However, this conditioning introduces correlations between rounds, and in particular it impacts our assertion that the ex-ante LW is at most $\rho_k$ per round.
We address this by explicitly bounding the impact of such correlations and arguing that they are small with high probability.  This introduces the additive error term in the theorem statement.\footnote{We note that similar issues of corrleation arise when analyzing the LW of the budget-pacing algorithm of~\citet{balseiro2019learning}; our solution is a variation of an idea due to~\cite{Gaitonde-itcs23}.}

Another technical issue that arises is specific to handling budget and ROI constraints simultaneously.  The intuition above does not carefully account for rounds in which $\mu_{k,t}$ switches from lying strictly below $\gamma_k-1$ to strictly above $\gamma_k-1$ or vice-versa.  It turns out that these transition rounds introduce error terms that can accumulate substantially; indeed, when we said above that the total spend over an interval is very close to $t \times \rho_k$, this approximation can be off by up to $\overlinev+\rho_k$ \emph{per round}, dominating our entire approximation.  We handle this by considering separately those rounds in which $\mu_{k,t}$ is very close to the boundary $\gamma_k - 1$, and directly relate the outcomes to what would occur precisely on the boundary itself.  The resulting error terms are yet another source of the additive error in the theorem statement.

\subsection{Individual Regret Guarantees}\label{sec: multi-cons-indi}


In this subsection, we consider the performance of an individual autobidder $k$ on its optimization problem when applying~\pref{alg:bg_roi_bud}. We abstract away the bids of other agents as supplied by a (potentially adaptive) adversary.
In particular, we do not assume that the other agents are controlled by any particular algorithm. Our regret bound holds w.r.t. a non-standard benchmark (\pref{def: pacing-multiplier}) that specializes to the standard benchmark (best fixed bid in hindsight) in stationary environments.

Since we focus on just agent $k$ throughout this subsection, we will drop the subscript $k$.  For each round $t$, we will write $x_t = x_t(\mu)$ and $p_t = p_t(\mu)$ for the allocation and payment if bidder $k$ picks multiplier $\mu$ in round $t$.  Note that these depend on the realized value of bidder $k$ as well as the bids of the other auction participants.  Recall also that $x_t$ and $p_t$ are both weakly non-increasing in $\mu$.
The expectations in this section are taken with respect to the randomness in value profiles of agent $k$ as well as the bids supplied by the adversary. Further, we define the following quantities:



\begin{definition}\label{def: Z-and-V}
For any round $t\in[T]$, define the expected budget expenditure $Z_t^B(\mu)$, the expected ROI expenditure $Z_t^R(\mu)$, the expected ROI gain $\rho_t(\mu)$, and the expected gained value $V_t(\mu)$ when the bidder chooses multiplier $\mu$. Letting $(x)^+=\max\{x,0\}$,
\begin{align*}
Z_t^B(\mu) \triangleq \mathbb{E}\sbr{p_t(\mu)}
&\quad\text{and}\quad
    Z_t^R(\mu) \triangleq \mathbb{E}\sbr{(\gamma\, p_t(\mu)-v_t\, x_t(\mu))^+},\\
\rho_t(\mu) \triangleq \mathbb{E}\sbr{(v_t\, x_t(\mu) - \gamma\, p_t(\mu))^+}
&\quad\text{and}\quad
    V_t(\mu) \triangleq \E \sbr{v_t\, x_t(\mu)}.
\end{align*}
\end{definition}

{Note that $Z_t^B(\mu)$ and $Z_t^R(\mu)-\rho_t(\mu)$ are both non-increasing functions when $\mu\geq 0$ and $\mu\in[0,\gamma-1]$,}
as is $V_t(\mu)$. Formally, we have the following lemma.
\begin{lemma}\label{lem:monotonicity}
    For any $t\in[T]$, $Z_t^{B}(\mu)$ is monotonically non-increasing for $\mu\geq 0$ and $Z_t^R(\mu)-\rho_t(\mu)$ is monotonically non-increasing for $\mu\in[0,\gamma-1]$.
\end{lemma}
Note also that $Z_t^R(\mu) - \rho_t(\mu)$ is precisely the expected value of $(\gamma p_t(\mu) - v_t x_t(\mu))$; $Z_t^R(\mu)$ captures the positive part of this random variable, and $\rho_t(\mu)$ captures the negative part.

We also assume Lipschitzness of allocations and payments with respect to $\mu$, which implies that $Z_t^B$, $Z_t^R$, $\rho_t$, and $V_t$ are all Lipschitz as well.  This can be interpreted as a requirement that the allocation and payment functions are sufficiently smooth as a function of an autobidder's bid.\footnote{\label{foot:Lip}We note that this assumption could be satisfied by adding $O(\lambda)$ noise to the multiplier selected by any given autobidder, at a loss of welfare proportional to $1/\lambda$.}

\begin{assumption}\label{assum:Lip}
    $Z_t^B(\mu)$, $Z_t^R(\mu)$, $V_t(\mu)$, $\rho_t(\mu)$, $t\in[T]$ are all $\lambda$-Lipschitz, 
    for some $\lambda\geq 0$.
\end{assumption}

We use a non-standard regret benchmark based on \emph{per-round pacing multipliers}.

\begin{definition}\label{def: pacing-multiplier}
For each round $t\in [T]$, we define a value-optimizing multiplier $\mu^*_t$, subject to the time-averaged constraints applied to this round. The formal definition is as follows:
%
\begin{description}
\setlength{\itemsep}{0pt}
\setlength{\parsep}{0pt}
\setlength{\topsep }{0pt}
\item[budget-pacing multiplier] $\mu_{t}^{B^*}$ is any $\mu \in [0,\frac{\overlinev}{\rho}-1]$ with $Z_t^B(\mu) = \rho$, or $0$ if no such $\mu$ exists.

\item[ROI-pacing multiplier] $\mu_{t}^{R^*}$ is any $\mu \in [0,\gamma-1]$ with $Z_{t}^R(\mu) = \rho_t(\mu)$, or $0$ if no such $\mu$ exists.

\item[pacing multiplier]
$\mu_{t}^* := \max\{\mu_{t}^{B^*}, \mu_{t}^{R^*}\} \geq 0$.
\end{description}
\end{definition}


\noindent Thus, our notion of regret is defined as follows:
$\Reg(T)\triangleq
    {\textstyle \sum_{t\in[T]}}\; V_t(\mu_t^*) - \sum_{t\in[T]}V_t(\mu_t).$

\begin{remark}
While not necessarily optimal globally, the pacing multipliers represent a reasonable goal for an online bidding algorithm in a complex, adversarial environment. Recall that vanishing-regret bounds with respect to the standard benchmark of the best-fixed-multiplier are impossible against an adversary, even for the special case of budget constraint only, due to the spend-or-save dilemma \citep{balseiro2019learning,AdvBwK-focs19}. Our benchmark side-steps this dilemma in an arguably natural way. Indeed, value-maximization subject to expected constraints in each round rules out the possibility of ``saving" the budget or using the saved budget later.
\end{remark}

Our main regret bound depends on the amount of drift in the environment, captured (in a fairly weak way) by the path-lengths of the per-round pacing multipliers:
\[
P_T^R := {\textstyle \sum_{t\in[T]}}
    \left|\mu_{t}^{R^*}-\mu_{t+1}^{R^*}\right|
\;\text{ and }\;
P_T^B := {\textstyle \sum_{t\in[T]}}
    \left|\mu_{t}^{B^*}-\mu_{t+1}^{B^*}\right|.
\]


\begin{theorem}
\label{thm:ind_roi_bud}
Fix any distribution over the values of agent $k$.
Posit Lipschitzness (\pref{assum:Lip}). \pref{alg:bg_roi_bud} with parameters $\eta_B=\frac{1}{\sqrt{T}\rho}$
and
$\eta_R=\frac{1}{\overlinev\,\sqrt{T}}$ guarantees that
\begin{align*}
    \E\sbr{\Reg(T)}\leq \order\rbr{(P_T^R+1)^{\nicefrac{1}{4}}\;\lambda^{\nicefrac{3}{4}}\;
    (\gamma+1) \cdot T^{\nicefrac{7}{8}} + (\overline{v}+\rho)^{2}\;\rho^{-1.5}\;
    \sqrt{\lambda(1+P_T^B)}\cdot T^{\nicefrac{3}{4}}}.
\end{align*}
When parameters $\lambda, \gamma, \overlinev, \rho$ are all constants, we have
\[ \E\sbr{\Reg(T)}\leq \order\rbr{(P_T^R+1)^{\nicefrac{1}{4}}\cdot T^{\nicefrac{7}{8}} + (P_T^B+1)^{\nicefrac{1}{2}}\cdot T^{\nicefrac{3}{4}}}.\]
\end{theorem}

\begin{remark}
This guarantee is non-trivial when the path-lengths $P_T^B$, $P_T^R$ are $O(T^{\alpha})$, $\alpha<\nicefrac12$.
\end{remark}

\begin{remark}
Our guarantees match \citet{Gaitonde-itcs23} when there is only budget constraint. Indeed, the $P_T^R+1$ term in \pref{thm:ind_roi_bud} can be improved to $P_T^R+| \mu_1^{R} -\mu_1^{R^*}|^2$ (see \pref{eq:roi_telescope}). The latter term becomes $0$ without the ROI constraint (because then $\mu_1^{R} := \gamma-1 = 0$ and $\mu_t^{R^*}=0$ for all rounds $t$). So, the $T^{\nicefrac{7}{8}}$ term vanishes from the regret bound and only the $T^{\nicefrac{3}{4}}$ term remains.
\end{remark}

Let us specialize this guarantee to the \emph{stationary-stochastic environment}, where in each round the competing bids and the agent's value are drawn independently from a fixed joint distribution. Then  the pathlengths
are $P_T^B = P_T^R = 0$. Moreover, our benchmark boils down to \emph{constraint-feasible multipliers}:
$\mu\geq 0$ such that
\[  {\textstyle \sum_{t\in[T]}}\;\E[Z_t^B(\mu)]\leq B
    \text{ and }
    {\textstyle \sum_{t\in[T]}}\;\E[Z_t^R(\mu)-\rho_t(\mu)]\leq 0.\]
Specifically, the pacing multiplier $\mu_{t}^*$ in each round $t$ is simply the constraint-pacing multiplier $\mu$ that maximizes the expected value $V(\cdot) := V_t(\cdot)$ (which is the same in all rounds).

\begin{corollary}\label{cor:statioary}
Consider the stationary-stochastic environment under Lipschitzness (\pref{assum:Lip}). Assume that parameters $(\lambda, \gamma, \overlinev, \rho)$ are all constants.
Let $\Pi$ be the set of all constraint-feasible multipliers.
%
Then~\pref{alg:bg_roi_bud} with the same parameters as in~\pref{thm:ind_roi_bud}
satisfies
\[ T\cdot {\textstyle \max_{\mu\in \Pi}} V(\mu)-
\textstyle \sum_{t=1}^T V(\mu_t) \leq \order\rbr{T^{\nicefrac{7}{8}}}.\]
\end{corollary}

\begin{remark}\label{rem:multiplayer-regret}
While \pref{thm:ind_roi_bud} applies to a general adversarial environment, we do not immediately obtain a meaningful corollary for the ``multi-player" environment when the other agents deploy the same algorithm.
Indeed, we can upper-bound the path-lengths $P_T^B$, $P_T^R$ as $O((\eta_B+\eta_R)T)$, but plugging this back into \pref{thm:ind_roi_bud} does not yield $o(T)$ regret. This issue arises (even) for the special case of budget-only constraint, as noted in \citet{Gaitonde-itcs23}. That said, we do get a non-trivial guarantee when all other agents use our algorithm with different parameters $\eta'_B$, $\eta'_R$ that are sufficiently small (provided that \pref{assum:Lip} holds, e.g., as per \pref{foot:Lip}).
\end{remark}



\xhdr{Proof intuition.}
The high-level idea is to define auxiliary stochastic convex functions $H^R_t$ and $H^B_t$ that achieve their minima at $\mu_t^{R^*}$ and $\mu_t^{B^*}$ resp., and interpret the update rules for $\mu_t^R$ and $\mu_t^B$ as applying stochastic gradient descent (SGD) w.r.t.  these auxiliary functions.  We then relate the difference in obtained value $V_t(\mu_t^*) - V_t(\mu_t)$ by the total loss in these auxiliary functions. Then, we'd ideally use facts about SGD to bound the total loss in value relative to the optimal benchmark.

Unfortunately, this approach runs into many technical problems.
%
The first problem is relatively straightforward.
Recall that we have both a budget and an ROI constraint, but only one multiplier is used in each round; this can be either the budget-multiplier $\mu_t^B$ or the ROI-multiplier $\mu_t^R$, whichever is larger.  To avoid having to reason about which multiplier is being followed each round, we will actually bound the sum of regret experienced for both multipliers.  We think of this as decomposing our experienced regret into the sum of two counterfactual regrets: one for the case where we have only the budget constraint and bid according to $\mu_t^B$, and one for the case where we have only the ROI constraint and bid according to $\mu_t^R$.  For each of these two cases, we can bound the total loss in value $V_t$ w.r.t.  the differences $|Z_t^B(\mu_t^B) - Z_t^B(\mu_t^{B^*})|$ and $|Z_t^R(\mu_t^R) - Z_t^R(\mu_t^{R^*})|$, resp., which we can then relate to corresponding differences in our auxiliary functions. 

This raises a more fundamental problem.  We'd like to argue that $\mu_t^B$ evolves according to SGD on our auxiliary function $H_t^B$, and similarly for $\mu_t^R$ and $H_t^R$.  However, since we only receive feedback w.r.t.  the larger multiplier, the smaller  multiplier ($\mu_t^B$ or $\mu_t^R$) may not be updated according to the gradient of its corresponding loss function.  One thing we do know is that, since the auxiliary functions are convex, the gradient we use to update the smaller multiplier can only be more negative, in expectation, than its ``correct'' gradient.  At first this seems like an unacceptable source of error; if gradients are too negative, then (for example) $\mu_t^B$ could drift arbitrarily far from $\mu_t^{B^*}$ in the negative direction.  However, we are saved by  \pref{lemma:ROI.budget.constraint.ex.post}: since we know that the budget and ROI constraints will necessarily be satisfied at the end of the $T$ rounds, our algorithm will not actually suffer any loss of value due to bids being too large (and hence, multipliers being too small).  We can therefore think of the evolution of one of the multipliers, say $\mu_t^B$, as following a variant of SGD in which an adversary can, at will, perturb any given update step to be more negative; but in exchange, we only suffer losses when $\mu_t^B > \mu_t^{B^*}$.  As it turns out, the usual analysis of SGD extends to this variant, so we can conclude that our total loss is not too large.

There are some additional technical challenges to handle as well.  Most notably, the auxiliary loss function for the ROI-multiplier is not convex in general but only convex when $\mu\in[0,\gamma-1]$.  This requires us to handle separately the case where the budget-multiplier is greater than $\gamma - 1$, and omit such rounds from our accounting of losses w.r.t.  the ROI constraint.  This complicates our definition of counterfactual regret for ROI, but it turns out that the aggregate loss can still be bounded with some additional effort.

\section{Proof of \pref{thm:liquid_welfare_roi_budget} (Liquid welfare)}


We prove the theorem in three steps.  We first show that with high probability, there is no significant correlation between the progression of our algorithm up to time $t$ and the ex-ante benchmark evaluated in future rounds.  We then condition on this event and bound the liquid welfare obtained on a per-instance basis. Finally, we take an expectation over realizations to obtain the desired bound on expected liquid welfare.

\xhdr{Step 1: Bounds on ex-ante Allocate Rules.}
W.l.o.g. we only consider allocation rules $\mathbf{y}$ such that
\begin{align}
\label{eq:ex.ante.budget}
    \mathbb{E}_{\mathbf{v} \sim F}
    \left[ \tfrac{1}{\gamma_k}\; y_k(\mathbf{v})\; v_{k} \right] \leq \rho_k.
\end{align}
This is because for any $y$ that violates this constraint, we can always decrease the allocation for agent $k$ without affecting $\overline{W}(\mathbf{y}, F)$.


We
make a stronger claim that \pref{eq:ex.ante.budget} holds for every round $t$ in which $\mu_{k,t} > \gamma_k - 1$.  To this end, we will show that, with high probability, the ex-ante optimal allocation rule $y$ does not generate significantly different outcomes in rounds where $\mu_{k,t} > \gamma_k - 1$ and rounds where $\mu_{k,t} \leq \gamma_k - 1$.

For each agent $k$, define the following quantity:
\begin{align}
    R_k(\mathbf{v}) \triangleq \sum_t
    \left[  \mathbbm{1}\{ \mu_{k, t} \leq \gamma_k - 1 \}\;
    \tfrac{1}{\gamma_k}\; y_k(\mathbf{v})\; v_{k, t} + \mathbbm{1}\{ \mu_{k, t} > \gamma_k - 1 \}\; \rho_k \right].
\end{align}
We can then use the theory of concentration of martingales to establish that the following bound on $R_k(\mathbf{v})$ holds with probability at least $1 - 1/(\overline{v} n T)^2$:
\begin{align}
\label{eq:martingale}
    R_k(\mathbf{v}) \leq \rho_k\cdot T + \overline{v} \sqrt{T \log (\overline{v} n T) }.
\end{align}
\pref{eq:martingale} is proved in~\cite{Gaitonde-itcs23}
(see \pref{lem:martingale_conc} in \pref{app:aux}).
Now, taking a union bound over all agents $k \in [n]$, with probability at least $1 - 1/(\overline{v} T)^2$, we have:
\begin{align}
\label{eq:good_val_real_def}
    R_k(\mathbf{v}) \leq \rho_k\cdot T + \overline{v} \sqrt{T \log (\overline{v} n T) }, \qquad \forall k \in [n]
\end{align}
We say that a value realization is ``good'' if it satisfies \pref{eq:good_val_real_def}.

\xhdr{Step 2: Liquid Welfare of ``Good'' Value Realizations. } Fix any ``good'' value profile realization $v$. For any advertiser $k$ whose liquid welfare is $B_k$, from \pref{eq:good_val_real_def}, we know that:
\begin{align}
\label{eq:budget.R.bound}
    W_k(\mathbf{v}) = B_k \geq R_k(\mathbf{v}) - \overline{v} \sqrt{T \log (\overline{v} n T) }.
\end{align}

Now we look at those agents $A \subseteq [n]$ for which the liquid welfare is strictly less than $B_k$:
$$W_k(\mathbf{v}) = \frac{1}{\gamma_k}
    {\textstyle \sum_{t=1}^T}\; x_{k, t}v_{k, t} < B_k.$$


Like in \pref{eq:budget.R.bound}, we again wish to bound $W_k(\mathbf{v})$ with respect to $R_k(\mathbf{v})$.  To that end, we will derive a bound on $W_k(\mathbf{v})$ that accounts for variation in $\mu_{k,t}$.
Let $\eta_k=\max\{\eta_{k,R}, \eta_{k,B}\}$.
For each round $t$, Let $S_t \subseteq A$ denote the agents for whom $\mu_{k, t} \leq \gamma_k-1$, and $T_t \subseteq S_t$ for the agents for whom $\gamma_k-1 - \eta_k(\overlinev+\rho_k) < \mu_{k, t} \leq \gamma_k-1$.  That is, $S_t$ are the agents bidding ``high enough,'' and $T_t$ are the agents in $S_t$ that are ``close to'' the threshold value $\gamma_k-1$ in round $t$.
\begin{lemma}
\label{lem:epochs}
The following inequality is guaranteed if all agents $k\in[n]$ apply~\pref{alg:bg_roi_bud}:
{\small
\begin{align}
\label{eq:ir.sum}
    \sum_{k \in A} \frac{1}{\gamma_k} \sum_{t=1}^{T} x_{k, t}v_{k, t} \geq \sum_{k \in A} \sum_{t = 1}^{T} \left[ \mathbbm{1} ( k \in S_t ) \frac{1}{\gamma_k} x_{k, t}v_{k, t} - \mathbbm{1} ( k \in T_t ) p_{k, t} + \mathbbm{1} (\mu_{k, t} > \gamma_k-1 )\rho_k \right]
\end{align}
}
\end{lemma}
\begin{proof}
Fix some agent $k \in A$.  Divide the time interval $[1,T]$ into intervals $(I_1, I_2, \dotsc)$ in the following manner: each interval $I = [t_1, t_2)$ is a minimal interval such that $\mu_{k, t_1} \leq \gamma_k - 1$ and $\mu_{k, t_2} \leq \gamma_k - 1$.  That is, $\mu_{k,t} > \gamma_k - 1$ for all $t_1 < t < t_2$. Note that according to~\pref{lem:no more than gamma}, we know that $\mu_{k,t}=\mu_{k,t}^B$ when $t\in(t_1,t_2)$.

 We wish to bound $\frac{1}{\gamma_k}\sum_{t \in I} x_{k,t}v_{k,t}$ for each such interval $I$.  Note that if $\mu_{k, t_1} \leq \gamma_k-1-\eta_k (\overline{v}+\rho_k)$, then we must have $t_2 = t_1 + 1$ (since $\mu_{k, t_1+1} \leq \mu_{t_1}+\eta_k(\overlinev+\rho_k) \leq \gamma_k -1$).  Thus, when $\mu_{k, t_1} \leq \gamma_k - 1 - \eta_k (\overlinev+\rho_k)$, we have $\frac{1}{\gamma_k} \sum_{t=t_1}^{t_2 - 1} x_{k, t}v_{k, t} = \frac{1}{\gamma_k} x_{k, t_1}v_{k, t_1}$.

On the other hand, if $\gamma_k - 1 - \eta_k(\overlinev+\rho_k) < \mu_{k,t_1} \leq \gamma_k - 1$, we have
\begin{align*}
    \gamma_k-1 <  \mu_{k, t_2-1} = \mu_{k, t_2-1}^B
    = \mu_{k, t_1}^B+\eta_k\sum_{\tau=t_1}^{t_2-2}(p_{k, \tau}-\rho_k)\leq \gamma_k-1+\eta_k\sum_{\tau=t_1}^{t_2-2}(p_{k, \tau}-\rho_k),
\end{align*}
which means that $\sum_{\tau = t_1}^{t_2 - 2}(p_{k,\tau}-\rho_k) \geq 0$.  Since $p_{k,t} \leq b_{k,t}x_{k,t} < \frac{1}{\gamma_k}x_{k,t}v_{k,t}$ for all $t_1 < t < t_2$, we can conclude that
\begin{align*}
\frac{1}{\gamma_k} \sum_{t=t_1}^{t_2 - 1} x_{k, t}v_{k, t} \geq \frac{1}{\gamma_k}x_{k,t_1}v_{k,t_1} + \sum_{t=t_1+1}^{t_2-1} p_{k,t}
\geq \frac{1}{\gamma_k}x_{k,t_1}v_{k,t_1} - p_{k,t_1} + (t_2-t_1-1)\rho_k.
\end{align*}
Summing over all time steps, we conclude that
\begin{align*}
    \frac{1}{\gamma_k} \sum_{t=1}^{T} x_{k, t}v_{k, t}
    & \geq \sum_{t = 1}^{T} \bigg[ \mathbbm{1} (\mu_{k, t} \leq \gamma_k -1 ) \frac{1}{\gamma_k} x_{k, t}v_{k, t}  \nonumber\\
    & \qquad  - \mathbbm{1} (\gamma_k-1-\eta_k(\overlinev+\rho_k) < \mu_{k, t} \leq \gamma_k -1 ) p_{k, t} + \mathbbm{1} (\mu_{k, t} > \gamma_k -1 )\rho_k \bigg].
\end{align*}
Summing this inequality over all agents yields \pref{eq:ir.sum}.
\end{proof}


Our next goal is to relate the terms in the right-hand side of \pref{eq:ir.sum} with the corresponding terms in $R_k(\mathbf{v})$.  Fix some round $t$.  We will focus on the first two terms in the expression inside the summation on the right hand side of \pref{eq:ir.sum}.  Consider the agents in $T_t$, which are the agents for whom $\gamma_k-1 - \eta_k(\overlinev+\rho_k) < \mu_{k, t} \leq \gamma_k-1$. We have
\begin{align}
    &\sum_{k \in T_t} \left(\frac{1}{\gamma_k} x_{k, t}v_{k, t} - p_{k, t}\right) \nonumber\\
    &
    = \sum_{k \in T_t}  \left(\frac{1}{\gamma_k} y_k(\mathbf{v}_t)v_{k, t} - \frac{1}{\gamma_k} v_{k, t}(y_k(\mathbf{v}_t) - x_{k, t}) - p_{k, t} \right) \nonumber \\
    & \geq \sum_{k \in T_t}  \left(\frac{1}{\gamma_k} y_k(\mathbf{v}_t)v_{k, t} - \frac{v_{k, t}}{1+\mu_{k, t}}(y_k(\mathbf{v}_t) - x_{k, t}) - \eta_k (\overline{v}+\rho_k)^2 - p_{k, t} \right)  \nonumber \\
     & = \sum_{k \in T_t} \frac{1}{\gamma_k} y_k(\mathbf{v}_t)v_{k, t} - \sum_{k \in T_t}\frac{v_{k, t}}{1+\mu_{k, t}}(y_k(\mathbf{v}_t) - x_{k, t}) - \sum_{k \in T_t}p_{k, t} - |T_t|\eta_k(\overlinev+\rho_k)^2\nonumber,
\end{align}
where the inequality follows from the definition of $T_t$: if $y_k(\mathbf{v}_t) \geq x_{k, t}$ we use that $\mu_{k, t} \leq \gamma_k-1$ and hence $\frac{1}{\gamma_k} \leq \frac{1}{\mu_{k, t}+1}$, whereas if $y_k(\mathbf{v}_t) < x_{k, t}$ we use that $\mu_{k, t} \geq \gamma_k-1-\eta_k(\overlinev+\rho_k)$ and hence $\frac{1}{\gamma_k} \geq \frac{1}{\mu_{k, t}+1} - \eta_k(\overlinev+\rho_k)$.

On the other hand, for agents in $S_t \setminus T_t$, we have
\begin{align}
    \sum_{k \in S_t \setminus T_t} \frac{1}{\gamma_k} x_{k, t}v_{k, t}
    &
    \geq \sum_{k \in S_t \setminus T_t}  \left(\frac{1}{\gamma_k} y_k(\mathbf{v}_t)v_{k, t} - \frac{1}{\gamma_k} v_{k, t}(y_k(\mathbf{v}_t) - x_{k, t})^+ \right) \nonumber \\
    & \geq \sum_{k \in S_t \setminus T_t}  \left(\frac{1}{\gamma_k} y_k(\mathbf{v}_t)v_{k, t} - \frac{v_{k, t}}{1+\mu_{k, t}}(y_k(\mathbf{v}_t) - x_{k, t})^+ \right) \nonumber \\
     & = \sum_{k \in S_t \setminus T_t} \frac{1}{\gamma_k} y_k(\mathbf{v}_t)v_{k, t} - \sum_{k \in S_t\backslash T_t}\frac{v_{k, t}}{1+\mu_{k, t}}(y_k(\mathbf{v}_t) - x_{k, t})^+ \nonumber.
\end{align}

Let $U_t$ be the set of all agents in $T_t$, plus the agents in $S_t \setminus T_t$ such that $x_{k, t} \leq y_k(\mathbf{v}_t)$.  Then, adding our two inequalities together gives
\begin{align*}
&\sum_{k \in S_t} \frac{1}{\gamma_k} x_{k, t}v_{k, t} - \sum_{k \in T_t}p_{k, t}\\
&\qquad \geq \sum_{k \in S_t} \frac{1}{\gamma_k} y_k(\mathbf{v}_t)v_{k, t} - \sum_{k \in U_t}\frac{v_{k, t}}{1+\mu_{k, t}}(y_k(\mathbf{v}_t) - x_{k, t}) - \sum_{k \in T_t}p_{k, t} - |T_t|\eta_k(\overlinev+\rho_k)^2.
\end{align*}

We wish to bound the term $\sum_{k \in U_t}\frac{v_{k, t}}{1+\mu_{k, t}}(y_k(\mathbf{v}_t) - x_{k, t})$ from the inequality above.  Note that this is the exactly the difference in declared value (i.e., bid) for $y_k(\mathbf{v}_t)$ and $x_{k, t}$ for agents in $U_t$.  It is here where we use the fact that the underlying auction is a core auction.  From the definition of a core auction, this difference in bids is at most the sum of payments of agents not in $U_t$. 
Therefore,

\begin{align}
\label{eq:ir.error}
\sum_{k \in S_t} \left(\frac{1}{\gamma_k} x_{k, t}v_{k, t}\right) - \sum_{k \in T_t}p_{k, t}
     & \geq \sum_{k \in S_t} \frac{1}{\gamma_k} y_k(\mathbf{v}_t)v_{k, t} - \sum_{k \not\in U_t}p_{k, t} - \sum_{k \in T_t}p_{k, t} - |T_t|\eta_k(\overlinev+\rho_k)^2 \nonumber\\
     & \geq \sum_{k \in S_t} \frac{1}{\gamma_k} y_k(\mathbf{v}_t)v_{k, t} - \sum_{k}p_{k, t} - |T_t|\eta_k(\overlinev+\rho_k)^2,
\end{align}
where in the second inequality we used the fact that $T_t \subseteq U_t$, so the two sums of over payments are over disjoint sets of agents.
Summing up~ \pref{eq:ir.error} over all rounds and substituting into \pref{eq:ir.sum} and using the definitions of $W_k(\mathbf{v})$ and $R_k(\mathbf{v})$, we conclude that
\[ \sum_{k \in A}W_k(\mathbf{v}) \geq \sum_{k \in A}R_k(\mathbf{v}) - \sum_t\sum_{k = 1}^n p_{k,t} - \eta_k(\overlinev+\rho_k)^2nT.\]
Summing over all agents $k\in[n]$, we have that for every ``good'' value realization $v$,
\begin{align}
\label{eq:LW.aggregate.bound}
\sum_k W_k(\mathbf{v}) \geq \sum_{k}R_k(\mathbf{v}) - \sum_t\sum_{k = 1}^n p_{k,t} - \overline{v}n \sqrt{T \log (\overline{v} n T)} - \eta_k(\overlinev+\rho_k)^2nT.
\end{align}

\xhdr{Step 3: Bounding Expected Liquid Welfare. } Recall from~\pref{obs:LW.revenue} that the total revenue collected over all rounds will never be greater than the liquid welfare of the allocation.  In other words, $\sum_t \sum_{k=1}^n p_{k,t} \leq \sum_k W_k(\mathbf{v})$.   We can therefore rearrange \pref{eq:LW.aggregate.bound} to conclude that
\[ 2 {\textstyle \sum_k}\; W_k(\mathbf{v})
\geq {\textstyle \sum_{k}}\; R_k(\mathbf{v}) - \overline{v}n \sqrt{T \log (\overline{v} n T)} - \eta_k(\overlinev+\rho_k)^2nT. \]
Taking expectations over $v$ and conditioning on the good event, we conclude that our expected liquid welfare is at least half of the expected optimal liquid welfare with an error term that grows at a rate of $\order(\overlinev n\sqrt{T\log(\overlinev nT)})$, as we take $\eta_{k} \leq \frac{\overlinev}{\overlinev+\rho_k}\sqrt{\frac{\log(\overlinev n T)}{T}}$. This completes the proof.

\section{Proof of \pref{thm:ind_roi_bud} (Regret)}
\label{sec: proof-ind-roi-bud}

We flesh out the intuition from \pref{sec: multi-cons-indi} but omit some details which are relegated to \pref{app:main}.

We begin by formalizing our interpretation of~\pref{alg:bg_roi_bud} as applying a form of SGD.
We construct auxiliary loss functions $H^B_t(\mu) = \rho\mu - \int_{0}^\mu Z_t^B(\tau)d\tau$ and $H^R_t(\mu) = \int_{0}^\mu \rho_t(\tau) - Z_t^R(\tau)d\tau$.

Based on~\pref{lem:no more than gamma}, we have the following lemma, which shows that if the ROI multiplier is larger than the budget multiplier, then the ROI-multiplier is updated by applying a SGD on function $H_t^R(\mu)$, and
otherwise the budget-multiplier is updated by applying SGD on function $H_t^B(\mu)$.

\begin{lemma}\label{lem: gradient-unbiased}
    \pref{alg:bg_roi_bud} guarantees that:
    \begin{itemize}
    \item If $\mu_{t}^{R}\geq \mu_{t}^{B}$, $\E\left[\gamma p_t(\mu_t)-v_t x_t(\mu_t)\right] = Z_t^R(\mu_t)-\rho_t(\mu_t)$.
    \item If $\mu_{t}^{R} < \mu_{t}^{B}$, $\E\left[p_{t}(\mu_t)-\rho\right] =  Z_t^B(\mu_t)-\rho$.
    \end{itemize}
\end{lemma}
\begin{proof}
    For ROI-multiplier, direct calculation shows that
    \begin{align*}
        \E\sbr{(\gamma p_t(\mu_{t})-v_t)x_t(\mu_t)}
        &= \E\sbr{(\gamma p_t(\mu_{t})-v_tx_t(\mu_t))^+} - \E\sbr{(v_tx_t(\mu_t)-\gamma p_t(\mu_{t}))^+} \\ 
        &= Z_t^R(\mu_t) - \rho_t(\mu_t).
    \end{align*}
    For budget-multiplier, direct calculation shows that $\E\left[x_{t}p_{t}(\mu_t)-\rho\right] =  Z_t^B(\mu_t)-\rho$. Combining the above two equations completes the proof.
\end{proof}

We next establish the convexity and Lipschitzness of $H_t^B(\mu)$ and $H_t^R(\mu)$.
\begin{lemma}\label{lem:cvx both}
    $H_t^R(\mu)$ is $(\gamma\cdot\bar{v})$-Lipschitz and convex when $\mu\in[0,\gamma-1]$ and $H_t^B(\mu)$ is $(\bar{v} + \rho)$-Lipschitz and convex in $\mu\geq 0$.
\end{lemma}
\begin{proof}
    The result for $H_t^B(\mu)$ is proven in~Lemma D.2 in~\cite{Gaitonde-itcs23}.
    For the function $H_t^R(\mu)$ we have
    \begin{align*}
        \nabla H_t^R(\mu) = \rho_t(\mu) - Z_t^R(\mu) = \E\left[v_tx_t(\mu)-\gamma p_t(\mu)\right],
    \end{align*}
    which we show is increasing over $\mu\in [0,\gamma-1]$ according to~\pref{lem:monotonicity}. In addition, we have for all $\mu\in[0,\gamma-1]$, it holds that
       $ |\nabla H_t^R(\mu)|\leq \max\{\gamma p_t(\mu), v_t\} \leq \gamma \bar{v}$.
\end{proof}





We proceed to prove \pref{thm:ind_roi_bud}. First, we decompose the overall regret into the regret with respect to budget-multiplier and ROI-multiplier as follows:
\begin{align}
    \Reg &= \sum_{t=1}^T\left(V_t(\mu_t^*) - V_t(\mu_t)\right) \nonumber\\
    &\leq \sum_{t=1}^T\left(V_t(\mu_t^*) - V_t(\mu_t)\right)\one\{\mu_t\geq \mu_{t}^*\} \tag{$V_t(\mu)$ is decreasing in $\mu$}\nonumber\\
    &\leq \sum_{t=1}^T\left[\left(V_t(\mu_t^*)-V_t(\mu_{t}^{B})\right)\one\{\mu_{t}^{B}\geq \max\{\mu_{t}^{R},\mu_{t}^*\}\right] \nonumber\\ 
    &\qquad\qquad+ \sum_{t=1}^T\left[\left(V_t(\mu_t^*)-V_t(\mu_{t}^{R})\right)\one\{\mu_{t}^{R}\geq \max\{\mu_{t}^{B},\mu_t^*\}\}\right] \nonumber\\
    &\leq \sum_{t=1}^T\left[\left(V_t(\mu_t^*)-V_t(\mu_{t}^{B})\right)\one\{\mu_{t}^{B}\geq \mu_{t}^*\}\right] + \sum_{t=1}^T\left[\left(V_t(\mu_t^*)-V_t(\mu_{t}^{R})\right)\one\{\mu_{t}^{R}\geq\mu_{t}^*, \mu_{t}^{B}\leq \gamma-1\}\right] \nonumber\\
    &\leq \sum_{t=1}^T\left[\left(V_t(\mu_{t}^{B^*})-V_t(\mu_{t}^{B})\right)\one\{E_t^B\}\right] + \sum_{t=1}^T\left[\left(V_t(\mu_{t}^{R^*})-V_t(\mu_{t}^{R})\right)\one\{E_t^R\}\right],\label{eqn:step-0-decompose} 
\end{align}

\noindent where $E_{t}^{B}$ is the event that $\mu_{t}^B\geq \mu_t^{B^*}$ and $E_{t}^{R}$ is the event that $\mu_t^R\geq \mu_t^{R^*}$ and $\mu_t^{B}\leq \gamma-1$. The third inequality holds because $\mu_{t}^{R}\le\gamma-1$ due to~\pref{lem:no more than gamma}, since  $\mu_{t}^{B}\leq \mu_{t}^{R}\leq \gamma-1$ in the second term. The fourth inequality holds because $V_t(\mu)$ is non-increasing in $\mu$.

We split the rest of the proof into four steps.


\xhdr{Step 1: Upper bounding the difference of $V_t$ by the difference of $Z_t^B$ and $Z_t^R-\rho_t$.}

Using the monotonicity of $p_t(\mu)$, we show that for ROI-multiplier and budget-multiplier,
\begin{align}
    \left(V_t(\mu_{t}^{R^*}) - V_t(\mu_{t}^{R})\right)\one\{E_{t}^{R}\}
    &\leq \left(\frac{\lambda}{\beta}\left(Z_t^R(\mu_{t}^{R^*}) -\rho_t(\mu_t^{R^*})- Z_t^R(\mu_{t}^{R})+\rho_t(\mu_t^{R})\right) + \beta\lambda\right)\one\{E_{t}^{R}\},\label{eqn:step-1-ROI-sketch}\\
    \left(V_t(\mu_{t}^{B^*}) - V_t(\mu_{t}^{B})\right) \one\{E_{t}^{B}\}
    &\leq \left(\frac{\overlinev+\rho}{\rho}(Z_t^B(\mu_{t}^{B^*}) - Z_t^B(\mu_{t}^{B}))\right)\one\{E_{t}^{B}\},\label{eqn:step-1-BUD-sketch}
\end{align}
for any $\beta>0$ to be fixed later. The proof is in \pref{app: multi-cons-indi} (\pref{lem: W-V-relation-ROI}).

\xhdr{Step 2: Upper bounding the difference of $Z_t^R$ ($Z_t^B$) by the difference of $H_t^R$ ($H_t^B$).}

Next, we relate $Z_t^R(\mu_{t}^{R^*})-Z_t^R(\mu_{t}^{R})$ with $H_t^R(\mu_{t}^{R^*})-H_t^R(\mu_{t}^{R})$. Direct calculation shows that
\begin{align*}
    \left(H_t^R(\mu_{t}^{R}) - H_t^R(\mu_{t}^{R^*})\right) & \one\{E_t^R\} = \int_{0}^{\mu_{t}^{R}-\mu_{t}^{R^*}}\left(\rho_t(\tau+\mu_{t}^{R^*})-Z_t^R(\tau+\mu_{t}^{R^*})\right)d\tau \one\{E_{t}^{R}\}.
\end{align*}
Note that $g^R(x) = \rho_t(x+\mu_{t}^{R^*})-Z_t^R(x+\mu_{t}^{R^*})$ is a non-decreasing function of $x$ when $x\in [0,\mu_{t}^{R}-\mu_{t}^{R^*}]$ according to~\pref{lem:cvx both}. Also, we have $g^R(0)\geq 0$. Applying a technical lemma from~\cite{Gaitonde-itcs23} (see \pref{lem: H-sqrt-lemma}) with $f(x)=g^R(x)-g^R(0)$, we know that 
\begin{align*}
    &\left[\left(\rho_t(\mu_{t}^{R}) - Z_t^R(\mu_{t}^{R})\right) - \left(\rho_t(\mu_{t}^{R^*}) - Z_t^R(\mu_{t}^{R^*}) \right)\right]\one\{E_t^R\} \\
    &\leq \sqrt{4\lambda\left(H_t^R(\mu_{t}^{R})-H_t^R(\mu_{t}^{R^*}) - (\mu_{t}^{R}-\mu_{t}^{R^*})(\rho_t(\mu_{t}^{R^*})-Z_t^R(\mu_{t}^{R^*}))\right)}\one\{E_t^R\}.
\end{align*}

\noindent When $\mu_{t}^{R}\geq \mu_{t}^{R^*}$, we know that $(\mu_{t}^{R}-\mu_{t}^{R^*})(\rho_t(\mu_{t}^{R^*})-Z_t^R(\mu_{t}^{R^*}))\geq 0$ and we can obtain
\begin{align}\label{eqn:step-2-ROI}
&\left(Z_t^R(\mu_{t}^{R^*}) - \rho_t(\mu_t^{R^*}) - Z_t^R(\mu_{t}^{R}) + \rho_t(\mu_t^R)\right) \one\{E_t^R\} \leq \sqrt{4\lambda(H_t^R(\mu_{t}^{R})-H_t^R(\mu_{t}^{R^*}))}\one\{E_t^R\}.
\end{align}

\noindent For budget-constraint multiplier $\mu_{t}^{B}$, we similarly define $g^B(x) = \rho - Z_t^B(\mu)$. Note that $g^B(x)$ is non-decreasing and $\lambda$-Lipschitz. Applying~\pref{lem: H-sqrt-lemma} on $g^B(x)-g^B(0)$, we also have
\begin{align}\label{eqn:step-2-BUD}
\left(Z_t^B(\mu_{t}^{B^*}) - Z_t^B(\mu_{t}^{B}) \right)\one\{E_t^B\} \leq \sqrt{2\lambda(H_t^B(\mu_{t}^{B})-H_t^B(\mu_{t}^{B^*}))}\one\{E_t^B\}.
\end{align}

\xhdr{Step 3-1: Upper bounding the regret with respect to $H^R_t$}

Now we analyze the term $\Big(H_t^R(\mu_{t}^{R}) - H_t^R(\mu_{t}^{R^*})\Big)\one\{E_t^R\}$ and $\Big(H_t^B(\mu_{t}^{R}) - H_t^B(\mu_{t}^{B^*})\Big)\one\{E_t^R\}$. For the first term, note that $(H_t^R(\mu_{t}^{R}) - H_t^R(\mu_{t}^{R^*}))\one\{E_{t}^{R}\}\leq \langle g_t^{R,R}, \mu_{t}^{R}-\mu_{t}^{R^*}\rangle\one\{E_{t}^{R}\}$ where $g_t^{R,R} = \nabla H_t^R(\mu_{t}^{R})$. This is because $\mu_{t}^{R^*}\leq \mu_{t}^{R}\leq \gamma-1$ and $H_t^R(\mu)$ is convex when $\mu\in[0,\gamma-1]$ according to~\pref{lem:cvx both}. According to~\pref{lem: gradient-unbiased}, if $\mu_{t}^R\geq \mu_t^B$, then $\mu_t^R$ is updated by a stochastic gradient with mean $g_t^{R,R}=\nabla H_t^R(\mu_t^R)$. However, note that $\mu_{t}^{R}$ may not be updated using its own stochastic gradient on $H_t^R(\mu)$, but may be updated by the gradient $g_t^{R,B} = \nabla H_t^R(\mu_{t}^{B})$ if $\mu_{t}^{B}\geq \mu_{t}^{R}$. However, using the convexity of $H_t^R(\mu)$ when $\mu\in [0,\gamma-1]$, we have $g_t^{R,B}\geq g_t^{R,R}$ as $\mu_{t}^{R}\leq \mu_{t}^{B}\leq \gamma-1$. Therefore, let $g_t^R$ be the gradient that is used to update $\mu_{t}^{R}$ and we have
\begin{align}
    \left(H_t^R(\mu_{t}^{R}) - H_t^R(\mu_{t}^{R^*})\right) \one\{E_t^R\} &\leq \inner{g_t^{R,R}, \mu_{t}^{R}-\mu_{t}^{R^*}}\one\{E_t^R\}\leq \inner{g_t^{R}, \mu_{t}^{R}-\mu_{t}^{R^*}}\one\{E_t^R\}, \label{eqn: gradient-devi}
\end{align} 
since when $\mu_{t}^{B}\geq \mu_{t}^{R}$, $g_t^R>g_t^{R,R}$ and $\mu_{t}^{R}\geq\mu_{t}^{R^*}$. Let $\hat{g}_t^{R}$ be the empirical gradient of $\mu_{t}^{R}$ at round $t$ with $\E[\hat{g}_t^R]=g_t^R$. Then, applying the analysis of online gradient descent,

\begin{align}
    &\E\left[\sum_{t=1}^T \left(H_t^R(\mu_{t}^{R}) - H_t^R(\mu_{t}^{R^*})\right)\one\{E_t^R\} \right] \leq \E\left[\inner{\hat{g}_t^R, \mu_{t}^{R}-\mu_{t}^{R^*}}\one\{E_t^R\}\right] \nonumber\\
    &\leq \sum_{t=1}^T\left(\frac{\vert\mu_{t}^{R}-\mu_{t}^{R^*}\vert^2-\vert\mu_{t+1}^{R}-\mu_{t+1}^{R^*}\vert^2}{2\eta_R} + \frac{\eta_R}{2} (\gamma+1)^2\overlinev^2 + \frac{(\gamma-1)\vert\mu_{t}^{R^*}-\mu_{t+1}^{R^*}\vert}{\eta_R}\right)\one\{E_t^R\}  \label{eqn: bias-stab-roi},
\end{align}

where the last inequality is because $\vert\hat{g}_t^R\vert\leq (\gamma+1)\bar{v}$ and
\begin{align*}
    \vert\mu_{t+1}^R-\mu_t^{R^*}\vert^2 &\leq \vert\mu_{t}^R-\eta_R\wh{g}_t^R - \mu_t^{R^*}\vert^2 = \vert\mu_t^R-\mu_t^{R^*}\vert^2 - 2\eta_R\inner{\hat{g}_t^R, \mu_{t}^{R}-\mu_{t}^{R^*}} + \eta_R^2\vert\wh{g}_t^R\vert^2,\\
    \vert\mu_{t+1}^R-\mu_t^{R^*}\vert^2
    &= \vert\mu_{t+1}^R-\mu_{t+1}^{R^*}\vert^2 + 2\inner{\mu_{t+1}^R-\mu_{t+1}^{R^*}, \mu_{t}^{R^*}-\mu_{t+1}^{R^*}} + \vert\mu_{t}^{R^*}-\mu_{t+1}^{R^*}\vert^2 \\
    &\geq \vert\mu_{t+1}^R-\mu_{t+1}^{R^*}\vert^2 - 2\vert\mu_{t+1}^R-\mu_{t+1}^{R^*}\vert\cdot\vert \mu_{t}^{R^*}-\mu_{t+1}^{R^*}\vert \\
    &\geq \vert\mu_{t+1}^R-\mu_{t+1}^{R^*}\vert^2 - 2(\gamma-1)\vert \mu_{t}^{R^*}-\mu_{t+1}^{R^*}\vert.
\end{align*}

While generally online gradient descent gives $\order(\sqrt{T})$ regret, the challenge in bounding the terms on the right hand side of \pref{eqn: bias-stab-roi} is that with the condition $\one\{E_t^R\}$, the term $\vert\mu_{t}^{R}-\mu_{t}^{R^*}\vert^2-\vert\mu_{t+1}^{R}-\mu_{t+1}^{R^*}\vert^2$ cannot be telescoped after summation. Fortunately, with a more careful analysis on the dynamic of $\mu_t^R$, we show that
\begin{align}\label{eq:roi_telescope}
    \sum_{t=1}^T\frac{\vert\mu_{t}^{R}-\mu_{t}^{R^*}\vert^2-\vert\mu_{t+1}^{R}-\mu_{t+1}^{R^*}\vert^2}{2\eta_R}\cdot\one\{E_t^R\} \leq \order\left(\frac{\vert\mu_1^R-\mu_1^{R^*}\vert^2+(\gamma+1)P_T^R}{\eta_R}+\eta_R(\gamma+1)^2\overlinev^2T\right).
\end{align} 
\noindent (The proof is in \pref{app: multi-cons-indi}.)
Combining \pref{eq:roi_telescope} and \pref{eqn: bias-stab-roi} with $\eta_R = \frac{1}{\sqrt{T}\overlinev}$ gives
\begin{align}
    \E\left[{\textstyle \sum_{t\in[T]}} \left(H_t^R(\mu_{t}^{R}) - H_t^R(\mu_{t}^{R^*})\right)\one\{E_{t}^{R}\} \right] &\leq \order\left(\frac{(\gamma+1)^2(P_T^R+1)}{\eta_R}+\eta_R(\gamma+1)^2\overlinev^2 T\right) \nonumber\\
    &=\order\left((\gamma+1)^2\overlinev(P_T^R+1)\sqrt{T}\right).\label{eqn:roi_regret_H}
\end{align} 

\xhdr{Step 3-2: Upper bounding the regret with respect to $H^B_t$}

Next, we consider $H_t^B(\mu_{t}^{B}) - H_t^B(\mu_{t}^{B^*})$. Let $g_t^{B,B}=\nabla H_t^B(\mu_{t}^{B})$ and $g_t^{B,R}=\nabla H_t^B(\mu_{t}^{R})$.
Similar to \textbf{Step 3-1}, because of the convexity of $H_t^B(\mu)$ when $\mu\in[0,+\infty)$, we know that if $\mu_{t}^{B}\leq \mu_{t}^{R}$, we have $g_t^{B,R}\geq g_t^{B,B}$. Therefore, we have the following inequality similar to~\pref{eqn: gradient-devi}:
\begin{align}
    &\left(H_t^B(\mu_{t}^{B}) - H_t^B(\mu_{t}^{B^*})\right)\one\{E_t^B\} \leq \inner{g_t^{B}, \mu_{t}^{B}-\mu_{t}^{B^*}}\one\{E_t^B\},\label{eqn: gradient-devi-budget}
\end{align}
where $g_t^B=\nabla H_t^B(\mu_t)$. Then, similar to~\pref{eqn: bias-stab-roi}, using the fact that $\mu_t^{B^*}\leq \frac{\overlinev}{\rho}$ for all $t$, we have
\begin{align}
    &\E\left[\sum_{t=1}^T \left(H_t^B(\mu_{t}^{B}) - H_t^B(\mu_{t}^{B^*})\right)\one\{E_t^B\} \right] \nonumber \\
    &\leq \sum_{t=1}^T\left(\frac{\vert\mu_{t}^{B}-\mu_{t}^{B^*}\vert^2-\vert\mu_{t+1}^{B}-\mu_{t+1}^{B^*}\vert^2}{2\eta_B} + \frac{\eta_B}{2} (\rho+\overlinev)^2 + \frac{\overlinev\vert\mu_{t}^{B^*}-\mu_{t+1}^{B^*}\vert}{\rho\eta_B}\right)\one\{E_t^B\} \label{eqn: bias-stab-budget},
\end{align}
where $(\rho+\bar{v})$ is a universal upper bound of the empirical gradient for $\mu_t^B$. Again, different from the standard online gradient descent analysis, the term $|\mu_t^B-\mu_t^{B^*}|^2-|\mu_{t+1}^B-\mu_{t+1}^{B^*}|^2$ cannot be telescoped and we prove (similarly to \pref{eq:roi_telescope}), that
\begin{align}\label{eq: budget_telescope}
\sum_{t=1}^T\frac{\vert\mu_{t}^{B}-\mu_{t}^{B^*}\vert^2-\vert\mu_{t+1}^{B}-\mu_{t+1}^{B^*}\vert^2}{2\eta_B}\cdot \one\{E_t^B\} \leq \order\left(\frac{|\mu_1^B-\mu_1^{B^*}|^2}{\eta_B} + \frac{\overlinev P_T^B}{\rho\eta_B} + \eta_B(\overlinev+\rho)^2T\right).
\end{align} 
\noindent (The proof is in \pref{app: multi-cons-indi}.)
Combining~\pref{eq: budget_telescope} and \pref{eqn: bias-stab-budget} with $\eta_B=\frac{1}{\sqrt{T}\rho}$ gives
\begin{align}\label{eqn:step-3-BUD}
\E\left[\sum_{t=1}^T \left(H_t^B(\mu_{t}^{B}) - H_t^R(\mu_{t}^{B^*})\right)\one\{E_t^B\} \right]\nonumber&\leq \order\left(\frac{\left|\mu_1^B-\mu_1^{B^*}\right|^2}{\eta_B}+\frac{\overlinev P_T^B}{\rho\eta_B} + \eta_B(\overlinev+\rho)^2T\right) \nonumber\\
&\leq \order\left(\frac{\overlinev^2}{\eta_B\rho^2}+\frac{\overlinev P_T^B}{\rho\eta_B} + \eta_B(\overlinev+\rho)^2T\right) \nonumber\\
&\leq \order\left(\frac{(\overlinev+\rho)^2}{\rho}(1+P_T^B)\sqrt{T}\right).
\end{align}


\xhdr{Step 4: Combining all the above analysis.} Finally, we combine~\pref{eqn:step-0-decompose},~\pref{eqn:step-1-ROI-sketch},~\pref{eqn:step-1-BUD-sketch},~\pref{eqn:step-2-ROI},~\pref{eqn:step-2-BUD},~\pref{eqn:roi_regret_H},~\pref{eqn:step-3-BUD} and obtain the following
\begin{align*}
    &\E\left[\sum_{t=1}^T\left(V_t(\mu_t^*)-V_t(\mu_t)\right)\right] \\
    &\leq  \E\left[\sum_{t=1}^T\left(\frac{\gamma}{\beta}\left(Z_t^R(\mu_{t}^{R^*}) - \rho_t(\mu_t^{R^*}) - Z_t^R(\mu_{t}^{R}) + \rho_t(\mu_{t}^{R}) \right) + \beta\lambda\right)\one\{E_t^R\}\right]   \\
    &\qquad + \E\left[\sum_{t=1}^T\frac{\overline{v}+\rho}{\rho}\left(Z_t^B(\mu_{t}^{B^*}) - Z_t^B(\mu_{t}^{B})\right)\one\{\mu_{t}^{B}\geq \mu_{t}^{B^*}\}\right] \tag{\pref{eqn:step-1-ROI-sketch},~\pref{eqn:step-1-BUD-sketch}}
    \\
    &\leq \E\left[\sum_{t=1}^T\left(\frac{\gamma}{\beta}\sqrt{4\lambda(H_t^R(\mu_t^R)-H_t^R(\mu_t^{R^*}))} + \beta\lambda\right)\one\{E_t^R\}\right] \\
    &\qquad + \E\left[\sum_{t=1}^T\frac{\overline{v}+\rho}{\rho}\sqrt{2\lambda(H_t^B(\mu_t^B)-H_t^B(\mu_t^{B^*}))}\one\{\mu_{t}^{B}\geq \mu_{t}^{B^*}\}\right] \tag{\pref{eqn:step-2-ROI},~\pref{eqn:step-2-BUD}}\\
    &\leq \E\left[\left(\frac{\gamma}{\beta}\sqrt{4T\sum_{t=1}^T\lambda(H_t^R(\mu_t^R)-H_t^R(\mu_t^{R^*}))} + \beta\lambda\right)\one\{E_t^R\}\right] \\
    &\qquad + \E\left[\frac{\overline{v}+\rho}{\rho}\sqrt{2T\lambda\sum_{t=1}^T(H_t^B(\mu_t^B)-H_t^B(\mu_t^{B^*}))}\one\{\mu_{t}^{B}\geq \mu_{t}^{B^*}\}\right] \tag{Jensen's inequality}\\
    &\leq \order\left(\frac{\gamma}{\beta}\sqrt{\lambda T^{1.5}(\gamma+1)^{2}\overlinev(P_T^R+1)} + \beta\lambda T + \frac{(\overline{v}+\rho)^{2}}{\rho^{1.5}}\sqrt{T^{1.5}\lambda(1+P_T^B)}\right)\tag{\pref{eqn:roi_regret_H},~\pref{eqn:step-3-BUD}} \\
    &\leq \order\left((P_T^R+1)^{\frac{1}{4}}\lambda^{\frac{3}{4}}(\gamma+1)T^{\frac{7}{8}} + (\overline{v}+\rho)^{2}\rho^{-1.5}\sqrt{\lambda(1+P_T^B)}T^{\frac{3}{4}}\right),
\end{align*}
\noindent where we obtain the last inequality is by picking the optimal $\beta>0$. This completes the proof.

\section{Numerical Experiments}
\label{sec:experiment}
We provide numerical experiments on several simulated problem instances with budget and ROI constraints, so as to illustrate our algorithm's performance in ``multi-player" environments when all advertisers run the same algorithm. In particular, individual regret in this environment is only studied via experiments, as we cannot meaningfully upper-bound it with our theory (see \pref{rem:multiplayer-regret} in \pref{sec: multi-cons-indi}). We also consider a few other algorithms, to get a sense of the problem difficulty and verify that our problem instances are indeed non-trivial. (In each experiment, all bidders are assigned the same algorithm.) All plots are deferred to \pref{app:expt-plots}.


\xhdr{Our algorithm(s).}
We consider a variant of \pref{alg:bg_roi_bud} with the budget-multiplier initialized as $\mu_{k,1}^B =\frac{1}{2\rho_k}$ instead of $\mu_{k,1}^B = \frac{\overlinev}{\rho_k}-1$, the latter being too conservative in practice. This variant enjoys the same aggregate and individual guarantees as \pref{alg:bg_roi_bud}, except that the budget constraint is violated by  $\leq O(\sqrt{T})$ with high probability. (All proofs carry over with minimal modifications.)

Moreover, we consider an \emph{optimistic variant} of our algorithm, in line with optimistic variants of online gradient descent (OGD) and online mirror descent (OMD) which have been prominent in recent prior work on repeated multi-agent games
\cite{syrgkanis2015fast,daskalakis2021near,NEURIPS2020_Golowich,Haipeng-iclr21-lastIterate}.
These variants have lead to provable guarantees in terms of regret and convergence in several scenarios when the original OGD and OMD do not appear to be amenable to analysis.%
\footnote{However, this prior work does not directly consider neither repeated auctions nor scenarios in which the players have global constraints, to the best of our knowledge.}
The general template for round $t$ of ``optimistic OGD" is as follows:
\begin{align*}
\mathtt{action}_t \leftarrow
    \mathtt{action}_t + \mathtt{update}_t + \rbr{\mathtt{update}_t - \mathtt{update}_{t-1}}.
\end{align*}
Accordingly, the optimistic variant of our algorithm updates the multipliers as
\begin{align}
\mu^R_{k, t+1}
    &= \mu^R_{k, t}
        + 2\,\eta_R\rbr{ x_{k,t}(\gamma_k\, p_{k,t} - v_{k,t})}
        - \eta_R\rbr{x_{k,t-1}(\gamma_k\, p_{k,t-1} - v_{k,t-1}) }, \\
\mu^B_{k, t+1}
    &= \mu^B_{k, t}
        + 2\,\eta_B \rbr{x_{k,t}\, p_{k,t} - \rho_k}
        - \eta_B \rbr{x_{k,t-1}\, p_{k,t-1} - \rho_k}.
\end{align}
This variant is a \emph{heuristic}, with no provable guarantees.

\xhdr{An algorithm from prior work.} We consider an algorithm from \citet{feng2022online}, which in turn is based on an algorithm from \citet{Balseiro-BestOfMany-Opre}
(for budget constraints only). We call this algorithm \emph{Best-Of-Many-Worlds} (\emph{BOMW}), following the title of \citet{Balseiro-BestOfMany-Opre}. The algorithm is based on dual mirror descent. The analysis focuses on the stationary-stochastic environment, achieving $O(\sqrt{T})$ regret with no constraint violations.%
\footnote{Specifically, we follow Algorithm 5.1 in \citet{feng2022online}. Algorithm 5.2 therein avoids constraint violations altogether due to an additional  warm-up phase, which appears less suitable to be multi-player environment.} No aggregate guarantees for this algorithm are provided, and no individual guarantees beyond the stationary-stochastic environment.

\xhdr{``Naive" baselines.}
We also consider two ``naive" baselines: the greedy algorithm and the epsilon-greedy algorithm. Both are standard ``templates" in multi-armed bandits. The greedy algorithm \emph{exploit} in each round, i.e., chooses the best action given the current observations. The epsilon-greedy algorithm explores uniformly with some fixed probability $\eps>0$, and exploits otherwise. In particular, the greedy algorithm ignores the need for exploration, and the epsilon-greedy algorithm (in our setting) ignores the non-stationarity of the multi-player environment and the constraints. So, we expect these algorithms to fail, and our experiments confirm this is indeed the case.


The two algorithms are implemented in the following unified way. For computational efficiency, we divide the rounds in batches of $M$ rounds each, where $M$ is a hyper-parameter, and update the multipliers only in the beginning of each such batch; we use $M=\lfloor\sqrt{T}\rfloor$.
In each round $t$ when we do update the multipliers, each algorithm calculates the budget- and ROI-pacing multipliers $\mu_{t}^{B^*}$, $\mu_{t}^{R^*}$ from \pref{def: pacing-multiplier} based on the observations collected so far,
\footnote{Specifically, we interpret the history as an ``empirical distribution" for the stationary-stochastic environment, and define the pacing multipliers w.r.t. this distribution.}
uses them to update the multipliers $\mu_{t}^B$, $\mu_{t}^R$ for budget and ROI constraint, resp., and outputs $\mu_t=\max\{\mu_{t}^B, \mu_{t}^R\}$.

The ``greedy" update sets $\mu_{t}^B$, $\mu_{t}^R$ to the resp. pacing multipliers:
    $\mu_{t}^{B} \leftarrow \mu_{t}^{B^*}$ and $\mu_{t}^{R} \leftarrow \mu_{t}^{R^*}$.

The epsilon-greedy update is as follows. With probability $1-\eps$, follow the greedy algorithm. Else, explore uniformly, namely: choose $\mu_{t}^B$ and $\mu_t^{R}$ independently and uniformly at random within their respective ranges: $\rbr{0,\frac{\overlinev}{\rho_k}-1}$ and $[0,\gamma-1]$. We use $\eps = 0.1$.

\xhdr{Problem Instances.}
We strived for a variety of problem instances, while keeping the experiments manageable. Recall that a problem instance in our model is defined by the per-round auction, the number of agents $K$, constraint parameters $\rho_k = B_k/T$ and $\gamma_k$ for each agent $k\in [K]$, and the distribution $F$ from which the value profiles are drawn. We endowed all players with the same constraint parameters: $\rho_k = \rho$ and $\gamma_k=\gamma$. We kept $K$ and $\gamma$ the same throughout all experiments. We considered the following variations: first-price or second-price auctions, two different values for $\rho$, and three different distributions $F$ (allowing both independence and correlation across bidders). Thus, our space of experiments is as follows:
$\{ \text{2 auctions}\}\times \{\text{high/low budgets}\} \times \{\text{3 choices for $F$}\}, $
for the total of $2\times 2\times 3 = 12$ choices. We represent them as $4\times 3$ matrix of plots, so that all experiments are laid out on the same page.%
\footnote{This desiderata to visualize all experiments on the same page was one reason to limit the space of experiments.} The numerical choices are as follows: $K=16$ (neither ``too small" nor ``too large"), $\rho\in \{0.15, 0.25\}$, $\gamma =1.5$, and time horizon $T=9000$.

Each problem instance is repeated $N=8$ times, with results averaged across the runs.

The three distributions $F$ that we consider are as follows:
\begin{OneLiners}
\item draw each value $v_k$ independently and uniformly at random from some interval, namely $[0,1]$.

\item draw each value $v_k$ independently from some fixed Gaussian distribution, clipping all values within range $[0,1]$. Namely, use mean $0.4$ and variance $0.2$.

\item draw the entire value profile from a correlated Gaussian distribution, clipping all values within range $[0,1]$. Namely, use mean $0.4$ for each agent, and covariance matrix $\Sigma=AA^\top$ where $A$ is a random matrix with each entry uniformly chosen from $[-0.5, 0.5]$.
    \footnote{Matrix $A$ is generated once for each of the $N=4$ runs, and then kept the same throughout all experiments. Hence, only 4 different matrices have been generated.}
\end{OneLiners}






\xhdr{Performance metrics} target individual and aggregate guarantees as well as convergence:

\begin{description}

\item[Multiplier dynamics]
    Focusing on agent $1$, we plot multiplier $\mu_{1,t}$ (averaged over runs) as a function of time $t$ (\pref{fig:mu}).

\item[Constraint slackness] Time-averaged slackness in, resp., budget and ROI  constraint, for agent $k$:
\begin{align}\label{eq:slack-expts}
 {\textstyle \frac{1}{T}\;\sum_{t=1}^T}\; \rbr{p_{k,t}-\rho}
\quad\text{and}\quad
{\textstyle\frac{1}{T}\sum_{t=1}^T}\; \rbr{x_{k,t}\,v_{k,t}-\gamma\, p_{k,t}}.
\end{align}
Positive value means no violation in the respective constraint. For each constraint, we plot the empirical CDF of the slackness over all agents and all runs. The results are in  \pref{fig:budget_1} and \pref{fig:budget_2} for the budget constraint, and in \pref{fig:roi_1} and \pref{fig:roi_2} for the ROI constraint.

\item[Liquid welfare] Time-averaged liquid welfare up to a given round $t$:
\begin{align}\label{eq:LW-expts}
 \overline{\LW}_t
 = {\textstyle \sum_{\text{agents } k\in[K]}}\;
    \min\cbr{\rho,\; \tfrac{1}{\gamma\,t}\;\;
    {\textstyle \sum_{\text{rounds }\tau\leq t}}\; x_{k,\tau}\, v_{k,\tau}}.
 \end{align}
We average $\overline{\LW}_t$ over the $N=4$ runs, and plot the result over time (\pref{fig:liquid_welfare}).

\item[Time-averaged Regret] Focus on agent $1$. We consider regret in terms of value received, with respect to the best-in-hindsight, call it \emph{time-averaged static regret}. In a formula:
\begin{align}\label{eq:regret-expts}
\mathtt{StaticReg}_1(T)
    := {\textstyle \sum_{t\in [T]}}\; v_t\, \rbr{x_t(\mu^*) - x_t(\mu_t)},
\end{align}
where $\mu^*$ is the smallest multiplier (i.e., the multiplier leading to largest bids) that does not violate the constraints given the realized history. We consider time-averaged static regret
    $\bar{R}(t) = \frac{1}{t}\, \mathtt{StaticReg}_1(t)$,
average it over the runs, and plot this average vs time $t$ (\pref{fig:individual_regret_normal}).
\end{description}

\begin{remark}
Eq.~\eqref{eq:regret-expts} is a standard notion of regret from adversarial bandits. As such, this is an important metric to consider for a multi-player environment, even though one cannot guarantee vanishing regret in the adversarial environment.
\footnote{Recall that in the adversarial environment, any algorithm suffers from an approximation ratio in the worst case, even for the special case of budget constraints \citep{balseiro2019learning}.}
Recall that our own regret bounds are relative to a different benchmark: pacing multipliers (\pref{def: pacing-multiplier}). However, it is unclear how to compute this benchmark in a computationally efficient way.

\end{remark}

\xhdr{Our findings.}
We make the following observations from the figures:
\begin{itemize}
    \item The optimistic variant of our algorithm performed indistinguishably from the original variant, as far as our plots are concerned. Therefore we omit it from all plots.

    \item Both our algorithm and BOMW are OK on the constraints. Our algorithm satisfies them exactly, while BOMW exhibits small constraint violations.

    \item Both naive baselines \emph{fail}, in that they exhibit large violations on at least one of the constraints in every experiment. (This is unsurprising, as discussed before.)

    \item The dynamics does not always converge to a stationary state, at least not within the (fairly large) timeframe that we considered, neither for our algorithm nor for BOMW.


    \item All algorithms achieve similar liquid welfare. Hence, the advantage of our algorithm (at least on these problem instances) lies in not violating the constraints.

    \item Time-averaged static regret $\bar{R}(t)$ decreases over time in a substantial way and appears to tend to $0$, across all experiments, both for our algorithms and for BOMW. The latter performs somewhat better as far as $\bar{R}(t)$ is concerned.

\end{itemize}

We zoom in on the static regret of our algorithm and fit it to the familiar shape $\mathtt{StaticReg}_1(T) = T^{\RegPow}$ for some $\RegPow>0$. To this end, \pref{fig:individual_regret} shows the log-log plot of for the cumulative static regret. We fit the curve by a linear function and show the corresponding slope and R-square value in \pref{tab:regression}, showing a strong empirical fit for 10 out of 12 instances. %
In the remaining two instances, which are both first-price auctions with $\rho = 0.25$, the static regret is non-monotone but small relative to the other cases. We omit these entries from \pref{tab:regression} due to the poor fit, but see \pref{fig:individual_regret} for a visualization.

\begin{table}[h]
    \centering
    \caption{The rate $\RegPow>0$ when we express the (cumulative) static regret of our algorithm as $T^{\RegPow}$. We obtain $\RegPow$ as a linear fit of the log-log curve. Each table entry is of the form ``$\RegPow$ (R-square value)''.}
    \begin{tabular}{cccc}
        \toprule
         & i.i.d Uniform   & i.i.d Gaussian   & Correlated Gaussian   \\
        \hline
        First-price, $\rho=0.15$  & 0.743 (0.9997) & 0.768 (0.9989)  & 0.739 (0.9992)       \\
        First-price, $\rho=0.25$  & / & 0.340 (0.9733)  & /        \\
        Second-price, $\rho=0.15$ & 0.799 (0.9996) & 0.811 (0.9994)  & 0.800 (0.9995)       \\
        Second-price, $\rho=0.25$ & 0.621 (0.9958) & 0.606 (0.9998)  & 0.668 (0.9981)       \\
        \bottomrule
    \end{tabular}
\label{tab:regression}
\end{table}

\xhdr{Main findings (summary).}
%
\emph{First}, our algorithm performs well on the metrics considered: no constraint violations, liquid-welfare performance same as BOMW, and vanishing static regret. Our problem instances appear non-trivial, given the performance of BOMW and the naive baselines.

\emph{Second}, the optimistic variant does not appear to improve performance of our algorithm. This is somewhat surprising, given its theoretical superiority in some other scenarios.

\emph{Third}, the dynamics does not always converge to a stationary state. This further motivates the study of aggregate guarantees such as ours that bypass convergence, and individual guarantees such as ours that go beyond the stationary environment. We observe strong algorithm performance regardless of whether the dynamics has converged.

\emph{Fourth}, our algorithm appears to have vanishing static regret, despite non-convergence of the dynamics. In fact, the observed regret fits well into the standard $T^{\RegPow}$ shape, for some fixed $\RegPow>0$ that we estimated empirically to be at most $0.82$ in all instances we simulated (and at most $0.75$ in most instances). We interpret this finding as further motivation for studying ``beyond the worst case" individual guarantees for multi-player environments.


\section{Conclusions and Open Problems}

We consider the problem of online bidding with both budget and ROI constraints, under a broad class of auction formats including first-price and second-price auction. {We set out to achieve both aggregate and individual guarantees, as expressed, resp., by liquid welfare and vanishing regret. We accomplish this with a novel variant of constraint-pacing, achieving
(i) the best possible guarantee in expected liquid welfare, (ii) vanishing individual regret against an adversary, and (iii) satisfying the budget and ROI constraints with probability $1$. The regret bound holds against a non-standard (albeit reasonable) benchmark, side-stepping impossibility results from prior work.}

Our work opens up several directions for future work. First, one would like to obtain similar results for other algorithms or classes thereof. Second, 
one would like to improve regret rates while maintaining a similar aggregate guarantee. This is a non-standard question for the literature on online bidding. Particularly interesting are regret bounds that go beyond a stationary stochastic environment (since the auction environment is often/typically not stationary in practice).
Another open direction is to analyze other aggregate market metrics such as platform revenue.  Here, it would be helpful to have a more complete understanding of the interaction between autobidders and tunable parameters like reserve prices.


\addcontentsline{toc}{section}{References}
\bibliography{refs,bib-abbrv,bib-AGT,bib-bandits,bib-slivkins,learning}

\begin{thebibliography}{38}
\providecommand{\natexlab}[1]{#1}
\providecommand{\url}[1]{\texttt{#1}}
\expandafter\ifx\csname urlstyle\endcsname\relax
  \providecommand{\doi}[1]{doi: #1}\else
  \providecommand{\doi}{doi: \begingroup \urlstyle{rm}\Url}\fi

\bibitem[Aggarwal et~al.(2019)Aggarwal, Badanidiyuru, and Mehta]{Gagan-wine19}
Gagan Aggarwal, Ashwinkumar Badanidiyuru, and Aranyak Mehta.
\newblock Autobidding with constraints.
\newblock In \emph{15th Workshop on Internet \& Network Economics (WINE)},
  pages 17--30, 2019.

\bibitem[Agrawal and Devanur(2019)]{AgrawalDevanur-ec14-OpRe}
Shipra Agrawal and Nikhil~R. Devanur.
\newblock Bandits with global convex constraints and objective.
\newblock \emph{Operations Research}, 67\penalty0 (5):\penalty0 1486--1502,
  2019.
\newblock Preliminary version in \emph{ACM EC 2014}.

\bibitem[Aumann(1974)]{Aumann-74}
Robert~J. Aumann.
\newblock Subjectivity and correlation in randomized strategies.
\newblock \emph{J. of Mathematical Economics}, 1:\penalty0 67–96, 1974.

\bibitem[Azar et~al.(2017)Azar, Feldman, Gravin, and Roytman]{Azar-wine17}
Yossi Azar, Michal Feldman, Nick Gravin, and Alan Roytman.
\newblock Liquid price of anarchy.
\newblock In \emph{10th Symp. on Algorithmic Game Theory (SAGT)}, pages 3--15,
  2017.

\bibitem[Babaioff et~al.(2021)Babaioff, Cole, Hartline, Immorlica, and
  Lucier]{Babaioff-itcs21}
Moshe Babaioff, Richard Cole, Jason~D. Hartline, Nicole Immorlica, and Brendan
  Lucier.
\newblock Non-quasi-linear agents in quasi-linear mechanisms (extended
  abstract).
\newblock In \emph{12th Innovations in Theoretical Computer Science Conf.
  (ITCS)}, pages 84:1--84:1, 2021.

\bibitem[Badanidiyuru et~al.(2018)Badanidiyuru, Kleinberg, and
  Slivkins]{BwK-focs13}
Ashwinkumar Badanidiyuru, Robert Kleinberg, and Aleksandrs Slivkins.
\newblock Bandits with knapsacks.
\newblock \emph{J. of the ACM}, 65\penalty0 (3):\penalty0 13:1--13:55, 2018.
\newblock Preliminary version in {\em FOCS 2013}.

\bibitem[Bailey and Piliouras(2018)]{Piliouras-ec18}
James~P. Bailey and Georgios Piliouras.
\newblock Multiplicative weights update in zero-sum games.
\newblock In \emph{ACM Conf. on Economics and Computation (ACM-EC)}, pages
  321--338, 2018.

\bibitem[Balseiro et~al.(2021)Balseiro, Deng, Mao, Mirrokni, and
  Zuo]{roi-balseiro2021robust}
Santiago Balseiro, Yuan Deng, Jieming Mao, Vahab Mirrokni, and Song Zuo.
\newblock Robust auction design in the auto-bidding world.
\newblock \emph{Advances in Neural Information Processing Systems}, 2021.

\bibitem[Balseiro and Gur(2019)]{balseiro2019learning}
Santiago~R Balseiro and Yonatan Gur.
\newblock Learning in repeated auctions with budgets: Regret minimization and
  equilibrium.
\newblock \emph{Management Science}, 65\penalty0 (9):\penalty0 3952--3968,
  2019.

\bibitem[Balseiro et~al.(2022{\natexlab{a}})Balseiro, Kroer, and
  Kumar]{StandardAuctions-ec22}
Santiago~R. Balseiro, Christian Kroer, and Rachitesh Kumar.
\newblock Contextual standard auctions with budgets: Revenue equivalence and
  efficiency guarantees.
\newblock In \emph{23rd ACM Conf. on Economics and Computation (ACM-EC)}, page
  476, 2022{\natexlab{a}}.

\bibitem[Balseiro et~al.(2022{\natexlab{b}})Balseiro, Lu, and
  Mirrokni]{Balseiro-BestOfMany-Opre}
Santiago~R. Balseiro, Haihao Lu, and Vahab~S. Mirrokni.
\newblock The best of many worlds: Dual mirror descent for online allocation
  problems.
\newblock \emph{Operations Research}, 2022{\natexlab{b}}.
\newblock Forthcoming. Preliminary version in \emph{ICML 2020}.

\bibitem[Borgs et~al.(2007)Borgs, Chayes, Immorlica, Jain, Etesami, and
  Mahdian]{Borgs-www07}
Christian Borgs, Jennifer~T. Chayes, Nicole Immorlica, Kamal Jain, Omid
  Etesami, and Mohammad Mahdian.
\newblock Dynamics of bid optimization in online advertisement auctions.
\newblock In \emph{16th Intl. World Wide Web Conf. (WWW)}, pages 531--540,
  2007.

\bibitem[Chen et~al.(2021)Chen, Kroer, and Kumar]{Chen-ec21}
Xi~Chen, Christian Kroer, and Rachitesh Kumar.
\newblock The complexity of pacing for second-price auctions.
\newblock In \emph{21st ACM Conf. on Economics and Computation (ACM-EC)}, page
  318, 2021.

\bibitem[Cheung and Piliouras(2019)]{Piliouras-colt19}
Yun~Kuen Cheung and Georgios Piliouras.
\newblock Vortices instead of equilibria in minmax optimization: Chaos and
  butterfly effects of online learning in zero-sum games.
\newblock In \emph{Conf. on Learning Theory (COLT)}, pages 807--834, 2019.

\bibitem[Conitzer et~al.(2018)Conitzer, Kroer, Sodomka, and
  Moses]{Conitzer-wine18}
Vincent Conitzer, Christian Kroer, Eric Sodomka, and Nicol{\'{a}}s E.~Stier
  Moses.
\newblock Multiplicative pacing equilibria in auction markets.
\newblock In \emph{14th Workshop on Internet \& Network Economics (WINE)}, page
  443, 2018.

\bibitem[Conitzer et~al.(2019)Conitzer, Kroer, Panigrahi, Schrijvers, Sodomka,
  Moses, and Wilkens]{Conitzer-ec19}
Vincent Conitzer, Christian Kroer, Debmalya Panigrahi, Okke Schrijvers, Eric
  Sodomka, Nicol{\'{a}}s E.~Stier Moses, and Chris Wilkens.
\newblock Pacing equilibrium in first-price auction markets.
\newblock In \emph{ACM Conf. on Economics and Computation (ACM-EC)}, page 587,
  2019.

\bibitem[Daskalakis and Panageas(2019)]{Daskalakis-itcs19-lastIterate}
Constantinos Daskalakis and Ioannis Panageas.
\newblock Last-iterate convergence: Zero-sum games and constrained min-max
  optimization.
\newblock In \emph{10th Innovations in Theoretical Computer Science Conf.
  (ITCS)}, volume 124, pages 27:1--27:18, 2019.

\bibitem[Daskalakis et~al.(2018)Daskalakis, Ilyas, Syrgkanis, and
  Zeng]{Daskalakis-iclr18}
Constantinos Daskalakis, Andrew Ilyas, Vasilis Syrgkanis, and Haoyang Zeng.
\newblock Training gans with optimism.
\newblock In \emph{6th International Conference on Learning Representations
  (ICLR)}, 2018.

\bibitem[Daskalakis et~al.(2021)Daskalakis, Fishelson, and
  Golowich]{daskalakis2021near}
Constantinos Daskalakis, Maxwell Fishelson, and Noah Golowich.
\newblock Near-optimal no-regret learning in general games.
\newblock \emph{Advances in Neural Information Processing Systems},
  34:\penalty0 27604--27616, 2021.

\bibitem[Dobzinski and Leme(2014)]{Shahar-icalp14}
Shahar Dobzinski and Renato~Paes Leme.
\newblock Efficiency guarantees in auctions with budgets.
\newblock In \emph{41st Intl. Colloquium on Automata, Languages and Programming
  (ICALP)}, pages 392--404, 2014.

\bibitem[Feng et~al.(2022)Feng, Padmanabhan, and Wang]{feng2022online}
Zhe Feng, Swati Padmanabhan, and Di~Wang.
\newblock Online bidding algorithms for return-on-spend constrained
  advertisers.
\newblock \emph{arXiv preprint arXiv:2208.13713}, 2022.

\bibitem[Gaitonde et~al.(2023)Gaitonde, Li, Light, Lucier, and
  Slivkins]{Gaitonde-itcs23}
Jason Gaitonde, Yingkai Li, Bar Light, Brendan Lucier, and Aleksandrs Slivkins.
\newblock Budget pacing in repeated auctions: Regret and efficiency without
  convergence.
\newblock In \emph{14th Innovations in Theoretical Computer Science Conf.
  (ITCS)}, 2023.

\bibitem[Gao et~al.(2022)Gao, Yang, Chen, Liu, and Karoui]{gao2022bidding}
Yuan Gao, Kaiyu Yang, Yuanlong Chen, Min Liu, and Noureddine~El Karoui.
\newblock Bidding agent design in the linkedin ad marketplace.
\newblock \emph{arXiv preprint arXiv:2202.12472}, 2022.

\bibitem[Golowich et~al.(2020{\natexlab{a}})Golowich, Pattathil, and
  Daskalakis]{Daskalakis-neurips20-lastIterate}
Noah Golowich, Sarath Pattathil, and Constantinos Daskalakis.
\newblock Tight last-iterate convergence rates for no-regret learning in
  multi-player games.
\newblock In \emph{33rd Advances in Neural Information Processing Systems
  (NeurIPS)}, 2020{\natexlab{a}}.

\bibitem[Golowich et~al.(2020{\natexlab{b}})Golowich, Pattathil, and
  Daskalakis]{NEURIPS2020_Golowich}
Noah Golowich, Sarath Pattathil, and Constantinos Daskalakis.
\newblock Tight last-iterate convergence rates for no-regret learning in
  multi-player games.
\newblock In \emph{Advances in Neural Information Processing Systems},
  volume~33, 2020{\natexlab{b}}.

\bibitem[Golrezaei et~al.(2021{\natexlab{a}})Golrezaei, Jaillet, Liang, and
  Mirrokni]{golrezaei2021bidding}
Negin Golrezaei, Patrick Jaillet, Jason Cheuk~Nam Liang, and Vahab Mirrokni.
\newblock Bidding and pricing in budget and roi constrained markets.
\newblock \emph{arXiv preprint arXiv:2107.07725}, 2021{\natexlab{a}}.

\bibitem[Golrezaei et~al.(2021{\natexlab{b}})Golrezaei, Lobel, and
  Paes~Leme]{roi-golrezaei2021auction}
Negin Golrezaei, Ilan Lobel, and Renato Paes~Leme.
\newblock Auction design for roi-constrained buyers.
\newblock In \emph{30th ACM Web Conf. (WWW)}, pages 3941--3952,
  2021{\natexlab{b}}.

\bibitem[Hart and Mas-Colell(2000)]{HartMasCollel-econometrica00}
Sergiu Hart and Andreu Mas-Colell.
\newblock A simple adaptive procedure leading to correlated equilibrium.
\newblock \emph{Econometrica}, 68:\penalty0 1127–1150, 2000.

\bibitem[Hazan(2015)]{Hazan-OCO-book}
Elad Hazan.
\newblock {Introduction to Online Convex Optimization}.
\newblock \emph{Foundations and Trends in Optimization}, 2\penalty0
  (3-4):\penalty0 157--325, 2015.
\newblock Published with \emph{Now Publishers} (Boston, MA, USA). Also
  available at {https://arxiv.org/abs/1909.05207}.

\bibitem[Immorlica et~al.(2022)Immorlica, Sankararaman, Schapire, and
  Slivkins]{AdvBwK-focs19}
Nicole Immorlica, Karthik~Abinav Sankararaman, Robert Schapire, and Aleksandrs
  Slivkins.
\newblock Adversarial bandits with knapsacks.
\newblock \emph{J. of the ACM}, August 2022.
\newblock Preliminary version in \emph{60th IEEE FOCS}, 2019.

\bibitem[Li and Tang(2022)]{roi-li2022auto}
Juncheng Li and Pingzhong Tang.
\newblock Auto-bidding equilibrium in roi-constrained online advertising
  markets.
\newblock \emph{arXiv preprint arXiv:2210.06107}, 2022.

\bibitem[Mehta and Perlroth(2023)]{roi-mehta2023auctions}
Aranyak Mehta and Andres Perlroth.
\newblock Auctions without commitment in the auto-bidding world.
\newblock \emph{arXiv preprint arXiv:2301.07312}, 2023.

\bibitem[Mertikopoulos et~al.(2018)Mertikopoulos, Papadimitriou, and
  Piliouras]{Piliouras-soda18}
Panayotis Mertikopoulos, Christos~H. Papadimitriou, and Georgios Piliouras.
\newblock Cycles in adversarial regularized learning.
\newblock In \emph{29th ACM-SIAM Symp. on Discrete Algorithms (SODA)}, pages
  2703--2717, 2018.

\bibitem[Moulin and Vial(1978)]{Moulin-78}
Herve Moulin and Jean-Paul Vial.
\newblock Strategically zero-sum games: the class of games whose completely
  mixed equilibria cannot be improved upon.
\newblock \emph{Intl. J. of Game Theory}, 7\penalty0 (3):\penalty0 201–221,
  1978.

\bibitem[Roughgarden et~al.(2017)Roughgarden, Syrgkanis, and
  Tardos]{DBLP:journals/jair/RoughgardenST17}
Tim Roughgarden, Vasilis Syrgkanis, and {\'{E}}va Tardos.
\newblock The price of anarchy in auctions.
\newblock \emph{J. Artif. Intell. Res.}, 59:\penalty0 59--101, 2017.

\bibitem[Slivkins(2019)]{slivkins-MABbook}
Aleksandrs Slivkins.
\newblock Introduction to multi-armed bandits.
\newblock \emph{Foundations and Trends$\circledR$ in Machine Learning},
  12\penalty0 (1-2):\penalty0 1--286, November 2019.
\newblock Published with \emph{Now Publishers} (Boston, MA, USA). Also
  available at {\tt https://arxiv.org/abs/1904.07272}.

\bibitem[Syrgkanis et~al.(2015)Syrgkanis, Agarwal, Luo, and
  Schapire]{syrgkanis2015fast}
Vasilis Syrgkanis, Alekh Agarwal, Haipeng Luo, and Robert~E Schapire.
\newblock Fast convergence of regularized learning in games.
\newblock \emph{Advances in Neural Information Processing Systems}, 28, 2015.

\bibitem[Wei et~al.(2021)Wei, Lee, Zhang, and Luo]{Haipeng-iclr21-lastIterate}
Chen{-}Yu Wei, Chung{-}Wei Lee, Mengxiao Zhang, and Haipeng Luo.
\newblock Linear last-iterate convergence in constrained saddle-point
  optimization.
\newblock In \emph{9th International Conference on Learning Representations
  (ICLR)}, 2021.

\end{thebibliography}



\appendix

\crefalias{section}{appendix} 

\section{Notation Summary}
\label{app: notation}
Let us list the notations used throughout the paper.\vspace{3mm}

\begin{tabular}{ll}
  $k\in [n]$ & agents \\
  $t\in [T]$ & rounds \\
  $X\in [0,1]^n$ & the set of feasible allocation profiles\\
  $\mathbf{x}_t\in X$ & allocation profile at round $t$, $\mathbf{x}_t=(x_{1,t},\dots,x_{n,t})$ \\
  $\mathbf{v}_t\in [0,\overlinev]^n$ & value profile at round $t$, $\mathbf{v}_t=(v_{1,t},\dots,v_{n,t})$ \\
  $\mathbf{b}_t$ & (effective) bid profile at round $t$, $\mathbf{b}_t=(b_{1,t},\dots,b_{n,t})$ \\
  $\mathbf{p}_t$ & payment profile at round $t$, $\mathbf{p}_t=(p_{1,t},\dots,p_{n,t})$ \\
  $\gamma_k$  & ROI constraint parameter for agent $k$ \\
  $B_k=\rho_k\cdot T$  & total budget for agent $k$ \\
  $W(\mathbf{x})$,~~$W_k(\mathbf{x})$
        & liquid welfare and agent $k$'s liquid value for allocation sequence $\mathbf{x}$ (\pref{def:liquid_welfare}) \\
  $\overline{W}(\mathbf{x},F)$  & ex-ante liquid welfare of allocation rule $\mathbf{x}$ and value distribution $F$ (\pref{def:liquid}) \\
  $\mu_{k,t}^B$,~~$\mu_{k,t}^R$
        & resp., budget-multiplier and ROI-multiplier for agent $k$ at round $t$ \\
  $\eta_{k,B}$,~~$\eta_{k,R}$  & learning rate for $\mu_{k,t}^B$ and $\mu_{k,t}^R$\\
  $Z_t^B$,~~$Z_t^R$ & expected budget- and ROI-expenditure at round $t$ (\pref{def: Z-and-V})\\
  $\rho_t$,~~$V_t$ & expected ROI and value gain at round $t$ (\pref{def: Z-and-V})\\
  $\mu_{t}^{B^*}$,~~$\mu_{t}^{R^*}$  & budget- and ROI-pacing multiplier at round $t$ (\pref{def: pacing-multiplier})\\
  $\mu_{t}^{*}$  & the pacing multiplier at round $t$ (\pref{def: pacing-multiplier})\\
  $P_{T}^{B}$ & path-length of budget-pacing multiplier, $P_{T}^{B}=\sum_{t=1}^{T-1}|\mu_t^{B^*}-\mu_{t+1}^{B^*}|$ \\
  $P_{T}^{R}$ & path-length of ROI-pacing multiplier, $P_{T}^{R}=\sum_{t=1}^{T-1}|\mu_t^{R^*}-\mu_{t+1}^{R^*}|$\\
\end{tabular}

\newpage
\section{Detailed Comparison to \citet{Gaitonde-itcs23} (Budgets Only)}
\label{app:comparison}

Our work builds upon results from \citet{Gaitonde-itcs23} on an autobidding 
with budget constraints but not ROI constraints.
Like~\cite{Gaitonde-itcs23}, our work also bounds the liquid welfare and individual regret obtained over the dynamics of an autobidding algorithm, and we employ a similar high-level strategy to obtain such bounds.   However, new technical challenges arise due to (a) fundamental differences between ROI constraints and budget constraints, and (b) extra complications due to handling multiple constraints simultaneously.  These challenges necessitate a new algorithm, changes to the way liquid welfare bounds are established, and a different methodology for establishing regret properties.  We elaborate on the similarities and differences below.

\xhdr{Algorithm.}
\citet{Gaitonde-itcs23} studied an existing {algorithm based on stochastic gradient descent (SGD)}, due to~\citet{balseiro2019learning} (henceforth BG). {This is our algorithm restricted to the special case of budget constraints only.}
{To handle multiple constraints we re-interpret the BG algorithm as maintaining a certain invariant whereby the choice of multiplier in round $t$ encodes the total slack in the budget constraint up to time $t$.  Our algorithm extends this interpretation to multiple constraints, which we view as distinct from standard multi-dimensional gradient descent.}
%
We emphasize that SGD-based autobidding algorithms for budget and ROI constraints already exist in the literature, as discussed in~\pref{sec:related}, but these algorithms do not appear amenable to our liquid welfare analysis.

\xhdr{Welfare analysis.}
Our liquid welfare analysis makes use of the fact that all constraints hold ex post with probability 1, since this enables us to compare revenue and liquid welfare.  This is easy to guarantee for budget constraints, but not for ROI.  Prior algorithms for ROI (that we are aware of) do not have this property,
and (as we discuss in~\pref{sec:just_ROI}) after just one round it is possible to reach a state from which the ROI constraint cannot be satisfied.  This challenge is further compounded with multiple constraints; a common strategy for multiple constraints is to estimate the likelihood each constraint will ultimately bind, but such methods tend to have some probability of failure.

Once we have established that all constraints are satisfied, the remainder of the liquid welfare analysis itself (Theorem 4.2) shares a high-level approach with \citep{Gaitonde-itcs23}, but enacts this approach differently.
The approach common to both papers is as follows: condition on the constraint that determines
the liquid welfare for a given agent, split rounds into intervals according to that agent's bid, and then bound the liquid welfare contributions from each interval 
separately.  These liquid welfare bounds work by charging any loss in welfare (relative to a benchmark) to revenue collected by the mechanism, either from the given agent or from other competing agents.
When doing this charging argument, it is important to take care when handling correlations introduced by the conditioning, and both papers handle this in a similar fashion.

We now discuss some differences.  One major distinction is that because we have multiple constraints, we must track which one to follow at any given round. However, this is not uniquely implied by the bids.  So our division into intervals does not perfectly correspond to the constraints to follow in our analysis.  We show that a certain bid threshold serves as an (imperfect on one side) proxy.  Unfortunately, whenever bids cross the threshold this generates errors that, by accumulating over rounds, have total magnitude that can be as large as the liquid welfare itself.  These errors would significantly degrade the approximation ratio.  To handle this problem we need a different accounting of the rounds, which makes it possible to use a charging argument for ROI constraints but introduces new additive errors that must be controlled.

\xhdr{Regret analysis.}
Finally, our new algorithm requires a new regret analysis.
The challenges and intuition are described in~\pref{sec: multi-cons-indi}; but we summarize them again here.

\emph{First,} since we only have (bandit) feedback for the larger multiplier, it is unclear why this provides a reasonable update for the smaller multiplier. We resolve this by noticing that the gradient we use to update the smaller multiplier can only be more negative than {the gradient we would have obtained by bidding with the smaller multiplier}, {and establishing loss bounds that are robust to these one-sided gradient errors.}

\emph{Second,} the auxiliary loss function for the ROI-multiplier is not convex in general, but only in certain parameter regimes.  We handle this by accounting for per-constraint losses in a way that lets us omit rounds and constraints where the ROI-multiplier may lie in a non-convex region.

\emph{Third,} since the algorithm transitions between budget-binding and ROI-binding time intervals, trivially adding up the regret in each interval leads to $\Theta(T)$ regret bound, 
since there can be $\Theta(T)$ intervals. We handle this using a careful analysis on the value of each multiplier at the beginning and the end of each interval.


\section{Details on Pacing Equilibria}
\label{app:pacing}
\noindent\textbf{An example for first-price auctions.}
We provide a simple example to lower-bound liquid welfare for first-price auctions (proving the statement in \pref{fn:pacing}).

\begin{claim}
Consider pacing equilibria in budget-constrained first-price auctions. For any $\lambda>\tfrac12$, there exists a pacing equilibrium whose liquid welfare does not exceed the $\lambda$ fraction of the optimum liquid welfare for the corresponding budget-constrained bidding game.
\end{claim}

\begin{proof}
Consider the following simple example. There is a single divisible item, one agent with (large) value $K$ per unit and budget $1$, and a second agent with value $1$ per unit and infinite budget. There is a pacing equilibrium where both agents bid $1$, and the first agent wins the entire item. This solution has liquid welfare $1$, whereas liquid welfare $2 - 1/K$ is possible: give the first agent $1/K$ of the item and the second agent $1 -1/K$ of the item.
\end{proof}

\noindent\textbf{Comparing the techniques.}
Our liquid welfare bound shares a common high-level proof strategy with the corresponding results for pacing equilibria (as well as with those for prophet inequalities and other "smooth pricing" methods), but with some important differences.

All these results stem from the following "approximate first-welfare theorem" approach: if an agent is obtaining much less (liquid) welfare at equilibrium than in an optimal allocation, this must be because they face high effective prices, which must be supported by some other agents’ payments. One can therefore charge any lost welfare against the total revenue collected by the platform, which is itself bounded by the (liquid) welfare. This proof strategy typically yields a 2- approximation. In a single-shot game, this charging argument can be applied directly to equilibrium bids.

The additional challenge in dynamic learning scenarios like ours is that, since bids are not necessarily in equilibrium, one cannot always charge lost welfare to high payments in the same round. To handle this, a now-standard idea is to amortize across rounds, by interpreting a low- regret sequence as a coarse correlated equilibrium (CCE). But unfortunately this doesn’t work in our setting: CCE is not strong enough to imply a constant PoA for liquid welfare; see \citet{Gaitonde-itcs23} for an example. So instead of appealing to low regret and CCE we analyze the dynamic sequence directly, separating ``high bids" (where the ROI constraint might bind) from ``low bids" (where only the budget constraint can bind) and using different amortization strategies for each. Many of our technical challenges come from setting up the learning algorithm to enable this analysis of the learning dynamics. 

\section{Auxiliary Lemmas from prior work}
\label{app:aux}

\begin{lemma}[Lemma 3.7 of~\cite{Gaitonde-itcs23}]\label{lem:martingale_conc}
Let $Y_1, \cdots, Y_T$ be random variables and $\mathcal{F}_0 \subseteq \cdots \subseteq \mathcal{F}_T$ be a filtration such that:
\begin{enumerate}
    \item $0 \leq Y_t \leq \overline{v}$ with probability 1 for some parameter $\overline{v} \geq 0$ for all $t$.
    \item $\mathbb{E}[Y_t] \leq \rho$ for some parameter $\rho$ for all $t$.
    \item For all $t$, $Y_t$ is $\mathcal{F}_t$-measurable but is independent of $\mathcal{F}_{t-1}$-measurable. Then:
    \begin{align*}
        \Pr\left({\textstyle \sum_{t\in[T]}}\; X_t Y_t + (1 - X_t)\rho \geq \rho \cdot T + \theta \right)
        \leq \exp \rbr{ -2 \theta^2/\rbr{T \overline{v}^2}}.
    \end{align*}
\end{enumerate}
\end{lemma}

\begin{lemma}[Lemma 4.12 of~\cite{Gaitonde-itcs23}]\label{lem: H-sqrt-lemma}
Let $f: \mathbb{R} \rightarrow \mathbb{R}$ be an increasing $\lambda-$Lipschitz function such that $f(0) = 0$. Let $R = \int_0^x f(y) dy$ for some $x \in \mathbb{R}$. Then $| f(x) | \leq \sqrt{2 \lambda R}$.
\end{lemma}

\begin{lemma}[Lemma B.1 of~\cite{Gaitonde-itcs23}]\label{lem: ex_ante_motiv}
Let 
$\Phi$ be an arbitrary allocation-sequence rule. 
Then there exists a (single-round) allocation rule ${{y}}: [0, \overline{v}]^n \rightarrow X$ such that
\begin{align*}
    \tilde{W}(\Phi, F) &:= \sum_{k=1}^n \min \left\{ B_k, \frac{1}{\gamma_k} \mathbb{E}_{\mathbf{v}\sim F} \left[ \sum_{t=1}^{T} \Phi_{k,t}(\mathbf{v})v_{k,t} \right] \right\} \\
    &= \sum_{k=1}^n T \cdot \min \left\{ \rho_k, \frac{1}{\gamma_k} \mathbb{E}_{\mathbf{v} \sim F}\left[ y_k(\mathbf{v}) v_k\right] \right\} = \overline{W}(\mathbf{y}, F).
\end{align*}
\end{lemma}

\newpage
\section{Warm-up proofs (\pref{sec:just_ROI})}
\label{app:just_ROI}

\subsubsection*{Proof of \pref{lem: warmup_1}: monotonicity of $v_tx_t(\mu)-\gamma p_t(\mu)$}

Let $d = \max_{j \neq k}b_j$, and let $\mu'$ be such that $v_t/(1+\mu') = d$.  Then for all $\mu > \mu'$ we have $x_t(\mu) = p_t(\mu) = 0$ (which is weakly increasing in $\mu$), and for all $\mu < \mu'$ we have that $x_t(\mu) = 1$ and $p_t(\mu)$ is weakly decreasing in $\mu$, so $v_t x_t(\mu) - \gamma p_t(\mu)$ is weakly increasing.

It only remains to establish what happens at the threshold $\mu = \mu'$, and then only when $\mu' \leq \gamma-1$.  Note however that when $\mu = \mu'$, the first and second highest bids are equal, so the payment of agent $k$ is determined to be $x_t(\mu) v_t/(1+\mu) \geq x_t(\mu) v_t/\gamma$.  This implies $v_t x_t(\mu') - \gamma p_t(\mu') \leq 0$, and hence $v_t x_t(\mu) - \gamma p_t(\mu) \leq 0$ for all $\mu < \mu'$ as well.  Since $v_t x_t(\mu) - \gamma p_t(\mu) = 0$ for all $\mu > \mu'$, we conclude that the difference is monotone in $\mu$ as claimed.


\subsubsection*{Proof of \pref{lemma:ROI.constraint.ex.post}: never violating the ROI constraint}
We prove this using induction. The base case follows trivially (since the multiplier is initialized to $\gamma - 1$). Now, suppose this is true for all time up to $t- 1$, i.e.,
\begin{align*}
    {\textstyle \sum_{\tau \in [t']}}\; v_{\tau} x_{\tau}
    \geq \gamma {\textstyle \sum_{\tau \in [t']}}\; p_{\tau}, \qquad \forall t' \leq t-1
\end{align*}
Consider time $t$. From the update rule, we have
\[\mu_t \geq \mu_0 + \eta\,\Lambda,
\quad \text{where } \Lambda :=
{\textstyle \sum_{\tau \in [t-1]}}\;
v_\tau x_\tau - \gamma p_\tau.\]


We consider two cases. First, suppose $\eta\,\Lambda<\gamma-1$.
Then
$ p_t \leq b_t \leq v_t/(\gamma - \eta\,\Lambda)$.
Using the fact that $\eta p_t \leq 1$ since $\eta < 1/\bar{v}$, we have: \[\gamma p_t + {\textstyle \sum_{\tau \in [t-1]}}\; p_\tau \leq v_t x_t + {\textstyle \sum_{\tau \in [t-1]}}\; v_\tau x_\tau,\]
which gives us the required claim. The second case is $\eta\,\Lambda
> \gamma - 1$. Then
$p_t \leq b_t \leq v_t$
which implies $\sum_{\tau \in [t]} v_\tau x_\tau - \gamma p_\tau > 0$. This completes the proof.

\section{Missing proofs from \pref{sec: multi-cons}}
\label{app:main}


\subsection{Proof of \pref{lemma:ROI.budget.constraint.ex.post}: never violating the constraints}
We first handle the ROI constraint, similarly to the proof of ~\pref{lemma:ROI.constraint.ex.post}.
We prove the following:
\begin{align} \label{eq:mainpf-ROI-induction}
{\textstyle \sum_{t\in[T]}}\; v_{k,t}x_{k,t}
\geq \gamma_k\; {\textstyle \sum_{t\in[T]}}\; p_{k,t}.
\end{align}
We use induction on $t$. We consider a fixed individual bidder and omit the subscript $k$ what follows. The base case still follows trivially, since the multiplier $\mu_{1}$ is initialized to $\max\{\gamma - 1, \frac{\overlinev}{\rho}-1\}$. Now, suppose this is true for all time up to $t-1$, i.e.,
\begin{align*}
    {\textstyle \sum_{\tau \in [t']}}\; v_{\tau} x_{\tau}
    \geq \gamma\; {\textstyle \sum_{\tau \in [t']}}\;
    p_{\tau}, \qquad \forall t' \leq t-1.
\end{align*}
Now, consider time $t$. From the update rule, we have $\mu_t^R \geq \mu_0^R + \eta_R \sum_{\tau = 1}^{t-1}\left(\gamma p_\tau - v_\tau x_\tau\right)$. We split the proof into two parts.

Suppose $\eta_R \sum_{\tau = 1}^{t-1}(v_\tau x_\tau - \gamma p_\tau) < \gamma -1$. This gives us:
\begin{align*}
p_t \leq b_tx_t = \frac{v_tx_t}{1+\mu_t} \leq \frac{v_tx_t}{1+\mu_t^R} \leq \frac{v_tx_t}{\gamma - \eta_R \sum_{\tau = 1}^{t-1}(v_\tau x_\tau - \gamma p_\tau)}.
\end{align*}
Now, using the fact that $\eta_R p_t \leq \eta_Rv_t\leq \eta\overlinev\leq 1$, we have:
\begin{align*}
    \gamma \sum_{\tau = 1}^{t} p_\tau =\gamma \left(p_t + \sum_{\tau = 1}^{t-1} p_\tau\right) \leq \eta_R\, p_t\sum_{\tau=1}^{t-1}(x_\tau v_\tau - \gamma p_\tau) + v_t x_t + \gamma \sum_{\tau = 1}^{t-1} p_\tau \leq \sum_{\tau=1}^tx_\tau v_\tau,
\end{align*}
where the last inequality uses the induction hypothesis.

Now, if $\eta_R \sum_{\tau = 1}^{t-1} (v_\tau x_\tau - \gamma p_\tau) > \gamma - 1$, we have $\sum_{\tau = 1}^{t-1} (v_\tau x_\tau - \gamma p_\tau)\geq (\gamma-1)\overlinev\geq (\gamma-1)v_tx_t$ and
$p_t \leq b_t \leq v_t$, which means that
\begin{align*}
    \gamma \sum_{\tau=1}^tp_\tau \leq \gamma \sum_{\tau=1}^{t-1}p_\tau + \gamma v_tx_t \leq \sum_{\tau=1}^tv_\tau x_\tau.
\end{align*}
Combining the above two claims finishes the proof for \pref{eq:mainpf-ROI-induction}.

For budget constraint, we prove that
    $\sum_{\tau\in[t]}\; p_{\tau}\leq \rho t$
using induction on the round $t$.
The base case holds as $p_1 \leq b_1 \leq \frac{v_1}{1+\mu_1^B}\leq \rho$. Suppose that the claim 
holds up to time $t-1$. According to the update rule, we know that $\mu_t^B\geq \mu_0^{B} + \eta_B\sum_{\tau=1}^{t-1}(x_\tau p_\tau-\rho)$. We also split the proof into two parts. Suppose $\eta_B\sum_{\tau=1}^{t-1}(x_\tau p_\tau - \rho) > 1-\frac{\overlinev}{\rho}$. Then we have
\begin{align*}
    p_t \leq b_tx_t = \frac{v_tx_t}{1+\mu_t} \leq \frac{v_tx_t}{1+\mu_t^B} \leq \frac{v_tx_t}{\eta_B\sum_{\tau=1}^{t-1}(x_\tau p_\tau - \rho) + \frac{\overlinev}{\rho}}.
\end{align*}
Therefore, we have:
\begin{align*}
    \sum_{\tau=1}^tp_\tau = \sum_{\tau=1}^{t-1}p_\tau + \frac{\rho}{\overlinev}\left(v_tx_t + \eta_Bp_t\sum_{\tau=1}^{t-1}(\rho- p_\tau)\right) \leq \sum_{\tau=1}^{t-1}p_\tau + \rho + \eta_B\rho\sum_{\tau=1}^{t-1}(\rho-p_\tau) = \rho t.
\end{align*}
If $\eta_B\sum_{\tau=1}^{t-1}(x_\tau p_\tau - \rho) < 1-\frac{\overlinev}{\rho}$, then we have $p_t\leq b_t\leq v_t$, which gives us
\begin{align*}
{\textstyle  \sum_{\tau\in[t]}}\; x_\tau p_\tau)
\leq \rho (t-1) + \overlinev \leq \rho t.
\end{align*}
Combining the above two inequalities finishes the proof.


\subsection{Proof of \pref{lem:no more than gamma}: $\mu_{t}^{R}\leq \gamma-1$}

    We prove this by induction. For conciseness, we omit the subscript of the agent index $k$. Base case trivially holds. Suppose that up to round $t$, $\mu_{t}^{R}\leq \gamma -1$ and $\mu_{t}^B\leq \frac{\overlinev}{\rho}-1$. First, consider $\mu_t^R$. At round $t+1$, if the bidder does not win an auction, then $\mu_{t+1}^R=\mu_{t}^{R}$. Otherwise, we have
    \begin{align*}
        \mu_{t+1}^R &= \mu_{t}^{R} + \eta_R\left(\gamma p_t(\mu_t)-v_t\right) \\
        & \leq \mu_{t}^{R} + \eta_R\left(\gamma p_t(\mu_{t}^{R})-v_t\right)
 \tag{$\mu_t=\max\{\mu_{t}^{B}, \mu_{t}^{R}, 0\}$ and $p_t(\mu)$ is non-increasing in $\mu$}\\
 &\leq \mu_{t}^{R} + \eta_R\left(\gamma\frac{v_t}{1+\mu_{t}^{R}}-v_t\right) \tag{payment does not exceed bid}\\
 &\leq \mu_{t}^{R} + \eta_R \frac{(\gamma  - 1 - \mu_{t}^{R})v_t}{1+\mu_{t}^{R}} \leq \mu_{t}^{R} + \gamma - 1 - \mu_{t}^{R}=\gamma - 1,
 \end{align*}
 where the last inequality uses the fact that $\eta_R\leq \frac{1}{\bar{v}}$. This proves the result for $\mu_t^R$.

Consider the budget-multiplier $\mu_t^B$. At round $t+1$, similarly, if the bidder does not win an auction, then $\mu_{t+1}^B = \mu_t^R-\eta_B\rho\leq \frac{\overlinev}{\rho}-1$. Otherwise, we have
\begin{align*}
    \mu_{t+1}^B &= \mu_t^B + \eta_B\left(p_t(\mu_t) - \rho\right) \\
    &\leq \mu_t^B + \eta_B\left(\frac{v_t}{1+\mu_t^B} - \rho\right) \tag{$p_t(\mu_t)\leq \frac{v_t}{1+\mu_t}\leq \frac{v_t}{1+\mu_t^B}$}\\
    &\leq 1+\mu_t^B + \frac{\eta_B\overlinev}{1+\mu_t^B} - \rho\eta_B - 1 \leq \frac{\overlinev}{\rho} + \frac{\eta_B\overlinev}{\overlinev/\rho} -\rho\eta_B-1 \leq \frac{\overlinev}{\rho}-1,
\end{align*}
where the third inequality is because $\eta_B\overlinev\leq 1$ and $h(x)=x+\frac{\eta_B\overlinev}{x}$ is increasing for $x\geq 1$.



\subsection{Proof of \pref{lem:monotonicity}: monotonicity}

    For any $t\in[T]$, since $p_t(\mu)$ is non-increasing in $\mu$, $Z_t^B(\mu)$ is non-increasing for $\mu\geq 0$. Next, consider $\rho_t(\mu)-Z_t^R(\mu)=\E[v_tx_t(\mu)-\gamma p_t(\mu)]$. Let $d_t$ denote the competing bid. Also note that the bid $b_t = v_t/(1 + \mu) > v_t / \gamma$. We split the proof into three cases:
    \begin{enumerate}
        \item If $d_t < v_t/\gamma$. We have $x_t(\mu) = 1$ (since $b_t > d_t$) and $p_t(\mu)$ is decreasing in $\mu$.
        \item If $v_t/\gamma \leq d_t \leq v_t$. Suppose the competing bid is $d_t = v_t/ (1 + \mu')$ for some constant $\mu' \in [1, \gamma-1]$. Then $x_t(\mu) = 1$ for $\mu \in [0, \mu'-1)$, and $v_t-\gamma p_t(\mu) < 0$ and increasing.

        At $\mu = \mu'$, we have $b_t = d_t$, and so $x_t(\mu) \leq 1$. Furthermore, we have $p_t(\mu) = x_t(\mu) v_t/(1 + \mu')$. Therefore $v_t x_t(\mu)-\gamma p_t(\mu) = x_t(\mu) \left (v_t - \gamma v_t/(1 + \mu') \right) \leq 0$, and also greater than $v_t x_t(\mu)-\gamma p_t(\mu)$,  $\mu < \mu'$ (since $p_t(\mu)$ decreases with $\mu$ and $x(\mu') \leq 1$). This shows that there will be an increase in $v_t x_t(\mu)-\gamma p_t(\mu)$ when we move from $\mu < \mu'$ to $\mu = \mu'$.

        Finally,  for $\mu \in (\mu', \gamma - 1]$, $v_t x_t(\mu)-\gamma p_t(\mu)$ will identically equal to zero. Therefore, we have $v_t x_t(\mu)-\gamma p_t(\mu)$ is increasing in this case.
        \item If $d_t > v_t$. In this case, since $b_t < d_t$, $x_t(\mu)$ and $p_t(\mu)$ will both be zero.
    \end{enumerate}
    Combining these three cases, and taking expectations gives the desired result.

\subsection{Proof of \pref{cor:statioary}: Regret in stationary-stochastic Environment}

Focus on the stationary-stochastic environment. Define $Z^B(\mu)\triangleq Z_t^B(\mu)$, $Z^R(\mu)\triangleq Z_t^R(\mu)$, $\rho(\mu)\triangleq\rho_t(\mu)$, and $V(\mu)\triangleq V_t(\mu)$ for all $t\in[T]$. Also, we have for all $t\in[T]$, $\mu_t^{R^*}=\mu^{R^*}$ and $\mu_t^{B^*}=\mu^{B^*}$. Here $\mu^{R^*}$ is any $\mu\in[0,\gamma-1]$ such that $\E[Z^R(\mu)-\rho(\mu)]=0$, or $0$ if no such $\mu$ exists; $\mu^{B^*}$ is any $\mu\in[0,\frac{\overlinev}{\rho}-1]$ such that $\E[Z^B(\mu)-\rho]=0$, or $0$ if no such $\mu$ exists.

    Now consider any $\mu\in\Pi$. As $\E[Z^B(\mu)]\leq \frac{B}{T}=\rho$ and $\E[Z^R(\mu)-\rho(\mu)]\leq 0$, according to the monotonicity of $Z^B(\mu)$ and $Z^R(\mu)-\rho(\mu)$ proven in~\pref{lem:monotonicity}, there exists $\mu^{B^*}\in[0,\frac{\overlinev}{\rho}-1]$ and $\mu^{R^*}\in[0, \gamma-1]$ such that $\mu\geq \mu^{B^*}$ and $\mu\geq \mu^{R^*}$. Therefore, according to~\pref{thm:ind_roi_bud} and the monotonicity of $V(\mu)$, we know that
    \begin{align*}
        {\textstyle \sum_{t\in [T]}}\; (V(\mu)-V(\mu_{t})) 
        \leq {\textstyle \sum_{t\in [T]}}\; (V(\max\{\mu_t^{B^*}, \mu_t^{R^*}\})-V(\mu_{t})) \leq \order(T^{\frac{7}{8}}),
    \end{align*}
    where the last inequality is because ~\pref{thm:ind_roi_bud} and $P_T^R=P_T^B=0$.

\section{Missing proofs from \pref{sec: proof-ind-roi-bud}}
\label{app: multi-cons-indi}




\begin{lemma}[Implies \pref{eqn:step-1-ROI-sketch}, \pref{eqn:step-1-BUD-sketch}]
\label{lem: W-V-relation-ROI}
For any $0\leq\mu_1\leq \mu_2\leq \gamma-1$ and any $\beta>0$, we have
\begin{align}\label{eq: W-V-relation-ROI-eq1}
    V_t(\mu_1) - V_t(\mu_2) &\leq \frac{\gamma}{\beta}\left(Z_t^R(\mu_1)-\rho_t(\mu_1)-Z_t^R(\mu_2)+\rho_t(\mu_2)\right) + \beta\lambda,
\end{align}
where $\lambda>0$ is the Lipschitz constant from \pref{assum:Lip}. Moreover, for any $0\leq\mu_1\leq \mu_2\leq \frac{\overline{v}}{\rho}-1$,
\begin{align}\label{eq: W-V-relation-ROI-eq2}
    V_t(\mu_1) - V_t(\mu_2) &\leq \frac{\overline{v}}{\rho}\left(Z_t^B(\mu_1)-Z_t^B(\mu_2)\right).
\end{align}
\end{lemma}

\noindent{\textbf{Proof }}
%
For any $\beta > 0$, as $V_t(\mu)$ is $\lambda$-Lipschitz based on~\pref{assum:Lip}, we have
\begin{align}
V_t(\mu_1) - V_t(\mu_2) &= V_t(\mu_1) - V_t(\min\{\mu_2, \gamma-1-\beta\}) + V_t(\min\{\mu_2, \gamma-1-\beta\}) - V_t(\mu_2)\nonumber\\
&\leq V_t(\mu_1) - V_t(\min\{\mu_2, \gamma-1-\beta\}) + \beta\lambda. \label{eqn:ir.add1}
\end{align}

Now we show that
\begin{align*}
&V_t(\mu_1) - V_t(\min\{\mu_2, \gamma-1-\beta\})\\ 
&\qquad\leq \frac{\gamma}{\beta}\left(Z_t^R(\mu_1) - \rho_t^R(\mu_1) - Z_t^R(\min\{\mu_2, \gamma-1-\beta\}) + \rho_t(\min\{\mu_2, \gamma-1-\beta\})\right).
\end{align*}
For any $\mu\in[\mu_1, \gamma-1-\beta]$,
\begin{align*}
    \nabla [Z_t^R(\mu)-\rho_t(\mu)]&= \nabla \mathbb{E}\left[((\gamma p_t(\mu)-v_t)x_t(\mu))\right] \\
    &= \mathbb{E}\left[\gamma \nabla p_t(\mu)x_t(\mu)\}\right] + \nabla\mathbb{E}\left[\gamma p_t(\mu) x_t(\mu)\}\right] - \nabla\E\left[v_t x_t(\mu)\right]\\
    &\leq \nabla\mathbb{E}\left[\gamma \frac{v_t}{1+\mu} x_t(\mu)\right] - \nabla V_t(\mu)\tag{$\nabla p_t(\mu)\leq 0$, $\nabla x_t(\mu)\leq 0$ and $p_t(\mu)\leq \frac{v_t}{1+\mu}$} \\
    & = \frac{\gamma - \mu - 1}{1+\mu}\nabla V_t(\mu).
\end{align*}

In addition, note that $Z_t^R(\mu_1)-\rho_t(\mu_1) - Z_t^R(\mu) + \rho_t(\mu) = \int_{\mu}^{\mu_1}\nabla (Z_t^R(\tau)-\rho_t(\tau))d\tau$ and $V_t(\mu_1) - V_t(\mu) = \int_{\mu}^{\mu_1}\nabla V_t(\tau)d\tau$. Therefore, we have
\begin{align*}
    &V_t(\mu_1) - V_t(\min\{\mu_2, \gamma-1-\beta\}) \\
    &\qquad= \int_{\mu_1}^{\min\{\mu_2, \gamma-1-\beta\}}-\nabla V_t(\tau)d\tau \\
    &\qquad\leq \frac{1+\min\{\mu_2, \gamma-1-\beta\}}{\gamma - \min\{\mu_2, \gamma-1-\beta\} - 1}\int_{\mu_1}^{\min\{\mu_2, \gamma-1-\beta\}}-\nabla (Z_t^R(\tau)-\rho_t(\mu))d\tau  \\
    &\qquad= \frac{1+\min\{\mu_2, \gamma-1-\beta\}}{\gamma - \min\{\mu_2, \gamma-1-\beta\} - 1}\\
    &\qquad\qquad\rbr{(Z_t^R(\mu_1) - Z_t^R(\min\{\mu_2, \gamma-1-\beta\} - \rho_t(\mu_1) + \rho_t(\min\{\mu_2, \gamma-1-\beta\})} \\
    &\qquad\leq \frac{\gamma}{\beta}\left(Z_t^R(\mu_1) -\rho_t(\mu_1) - Z_t^R(\mu_2)+\rho_t(\mu_2)\right). \tag{$Z_t^R(\mu)-\rho_t(\mu)$ is non-increasing in $\mu\in[0,\gamma-1]$}
\end{align*}




Plugging the above into \pref{eqn:ir.add1} gives
\begin{align}
\label{eqn:ir.value.ineq}
V_t(\mu_1) - V_t(\mu_2) \leq \frac{\gamma}{\beta} \left(Z_t^R(\mu_1) -\rho_t(\mu_1) - Z_t^R(\mu_2)+\rho_t(\mu_2)\right) + \beta\lambda,
\end{align}
which finishes the proof of \pref{eq: W-V-relation-ROI-eq1}. For \pref{eq: W-V-relation-ROI-eq2}, for any $\mu\in[0,\frac{\overline{v}}{\rho}-1]$ we have
\begin{align*}
\nabla Z_t^B(\mu) &= \nabla \mathbb{E}
\left[p_t(\mu)x_t(\mu)\right] \\
&=  \mathbb{E}
\left[\nabla p_t(\mu)x_t(\mu)\right] + \mathbb{E}
\left[ p_t(\mu) \nabla x_t(\mu)\right]\\
&\leq \nabla\mathbb{E}
\left[ \frac{v_t}{1+\mu}  x_t(\mu)\right] \tag{$\nabla p_t(\mu)\leq 0$, $\nabla x_t(\mu)\leq 0$ and $p_t(\mu)\leq \frac{v_t}{1+\mu}$} \\
& =\frac{1}{1+\mu}\;\nabla V_t(\mu)
\leq \frac{\rho}{\overline{v}}\; \nabla V_t(\mu).\\
V_t(\mu_1) - V_t(\mu_2)
&= \int_{\mu_1}^{\mu_2}-\nabla V_t(\tau)d\tau \leq \frac{\overlinev}{\rho}\int_{\mu_1}^{\mu_2}-\nabla Z_t^B(\tau)d\tau = \frac{\overlinev}{\rho}\left(Z_t^B(\mu_1) - Z_t^R(\mu_2)\right)
    \\
    &\leq \frac{\overlinev}{\rho}\left(Z_t^B(\mu_1) - Z_t^B(\mu_2)\right).
    \qquad\qquad\qed
\end{align*}

\noindent\textbf{Proof of \pref{eq:roi_telescope}.}
We decompose the $T$ rounds into $S$ time intervals $I_1=[1,e_1],\dots,I_S=[w_S, e_S]$, where each time interval is a maximal sequence of consecutive rounds such that $\mu_{t}^{R}\geq \mu_{t}^{R^*}$ and $\mu_{t}^{B}\leq \gamma-1$. Then we have
\begin{align}
    &\sum_{t=1}^T\frac{\vert\mu_{t}^{R}-\mu_{t}^{R^*}\vert^2-\vert\mu_{t+1}^{R}-\mu_{t+1}^{R^*}\vert^2}{2\eta_R}\cdot\one\{E_t^R\} \leq \sum_{s=1}^S\frac{\vert\mu_{w_s}^{R}-\mu_{w_s}^{R^*}\vert^2-\vert\mu_{e_s+1}^{R}-\mu_{e_s+1}^{R^*}\vert^2}{2\eta_R}  .\label{eqn: main-reg-roi}
\end{align}
For $s\geq 2$, consider the most recent round $\sigma_s$ before round $w_s$ such that $\mu_{\sigma_s}^{B}\leq \gamma-1$. As $\mu_{t}^{R}\leq \gamma-1$, we know that $\mu_{t}^{B}\geq \mu_{t}^{R}$ when $t\in[\sigma_s+1, w_s-1]$. In addition, according to the update rule of $\mu_{t}^{R}$, we know that when $t\in[\sigma_s+1, w_s-1]$ we have
\[\mu_{t+1}^{R} = \mu_{t}^{R} + \eta_R\left(\gamma p_t(\mu_t)-v_t\right)x_t \leq \mu_{t}^{R} + \eta_R \left(\frac{\gamma v_t}{1+(\gamma - 1)}\right)x_t \leq \mu_{t}^{R}.\]
Next, consider the round $\sigma_s$. If $\sigma_s$ belongs to some interval $I_i$, according to the definition of $\sigma_s$, $\sigma_s$ must be the end of $I_{s-1}$ (i.e. $\sigma_s$ = $e_{s-1}$). In this case, we have
\begin{align*}
    \vert\mu_{w_s}^{R}-\mu_{w_s}^{R^*}\vert^2 &\leq \vert\mu_{\sigma_s+1}^{R}-\mu_{w_s}^{R^*}\vert^2 \tag{$\mu_{w_s}^{R^*}\leq \mu_{w_s}^R\leq \mu_{\sigma_s+1}^R$}\\
    &= \vert\mu_{\sigma_s+1}^{R}-\mu_{\sigma_s+1}^{R^*}\vert^2 +2(\mu_{\sigma_s+1}^{R}-\mu_{\sigma_s+1}^{R^*})(\mu_{\sigma_s+1}^{R^*} - \mu_{w_s}^{R^*}) + \vert\mu_{\sigma_s+1}^{R^*}-\mu_{w_s}^{R^*}\vert^2 \\
    &\leq \vert\mu_{e_{s-1}+1}^{R}-\mu_{e_{s-1}+1}^{R^*}\vert^2 +3(\gamma-1)\Bigg(\sum_{t\in[e_{s-1}+1, w_s-1]}\vert\mu_{t}^{R^*}-\mu_{t+1}^{R^*}\vert\Bigg).
\end{align*}
Otherwise, $\sigma_s$ is outside the interval and $\mu_{\sigma_s}^{R}<\mu_{\sigma_s}^{R^*}$. From the update of $\mu_{t}^{R}$, we know that
$\mu_{\sigma_s+1}^{R} \leq \mu_{\sigma_s}^{R}+\eta_R(\gamma+1)\overlinev < \mu_{\sigma_s}^{R^*}+\eta_R (\gamma+1)\overlinev.$ Therefore we know that
\begin{align*}
    \vert\mu_{w_s}^{R}-\mu_{w_s}^{R^*}\vert^2 &\leq \vert\mu_{\sigma_s+1}^{R}-\mu_{w_s}^{R^*}\vert^2 \\&\leq \vert\mu_{\sigma_s}^{R^*}+\eta_R(\gamma+1)\overlinev-\mu_{w_s}^{R^*}\vert^2 \\
    &\leq \vert\mu_{\sigma_s}^{R^*}-\mu_{w_s}^{R^*}\vert^2 +2\eta_R(\gamma+1)\overlinev\cdot\sum_{\tau\in[\sigma_s,w_s-1]}\vert\mu_{\tau}^{R^*}-\mu_{\tau+1}^{R^*}\vert + \eta_R^2(\gamma+1)^2\overlinev^2 \\
    &\leq (\gamma-1)\vert\mu_{\sigma_s}^{R^*}-\mu_{w_s}^{R^*}\vert+2\eta_R(\gamma+1)\overlinev\cdot\sum_{\tau\in[\sigma_s,w_s-1]}\vert\mu_{\tau}^{R^*}-\mu_{\tau+1}^{R^*}\vert + \eta_R^2(\gamma+1)^2\overlinev^2 \\
    &\leq \order\Bigg((\gamma+1)\sum_{\tau\in [\sigma_s,w_s-1]}\left|\mu_\tau^{R^*}-\mu_{\tau+1}^{R^*}\right| + \eta_R^2(\gamma+1)^2\overlinev^2\Bigg).
\end{align*}
Combining the above two cases and noticing that in the second case, $\sigma_s$ does not belong to an interval $I_i$, we have for any $\eta_R\leq \min\{1,\frac{1}{\overlinev}\}$,
\begin{align*}
    &\sum_{s=1}^S\left(\frac{\vert\mu_{w_s}^{R}-\mu_{w_s}^{R^*}\vert^2-\vert\mu_{e_s+1}^{R}-\mu_{e_s+1}^{R^*}\vert^2}{\eta_R} \right) \nonumber\\
    &\leq \frac{\left|\mu_1^R-\mu_1^{R^*}\right|^2}{\eta_R}+\sum_{s=1}^{S-1}\left(\frac{\vert\mu_{w_{s+1}}^{R}-\mu_{w_{s+1}}^{R^*}\vert^2-\vert\mu_{e_s+1}^{R}-\mu_{e_s+1}^{R^*}\vert^2}{\eta_R} \right) \nonumber \\
    &\leq \frac{\left|\mu_1^R-\mu_1^{R^*}\right|^2}{\eta_R}+\sum_{s=1}^{S-1}\left(\frac{(\gamma+1)\sum_{\tau=e_s+1}^{w_{s+1}}\left|\mu_\tau^{R^*}-\mu_{\tau+1}^{R^*}\right|}{\eta_R}+\eta_R(\gamma+1)^2\overlinev^2 \right) \nonumber \\
    &\leq \order\left(\frac{\left|\mu_1^R-\mu_1^{R^*}\right|^2+(\gamma+1)P_T^R}{\eta_R} + \eta_R(\gamma+1)^2\overlinev^2T\right).
    \qquad\qquad\qed
\end{align*}


\xhdr{Proof of \pref{eq: budget_telescope}.}
Similar to the proof of \pref{eq:roi_telescope}, we decompose the total horizon $[T]$, into $S_b$ intervals $I_1=[1,e_1'],\dots,I_{S_b}'=[w_{S_b}', e_{S_b}']$, where each interval $I$ contains a maximal sequence of consecutive rounds such that $\mu_{t}^{B}\geq \mu_{t}^{B^*}$. Then we have
\begin{align}
    &\sum_{t=1}^T\frac{\vert\mu_{t}^{B}-\mu_{t}^{B^*}\vert^2-\vert\mu_{t+1}^{B}-\mu_{t+1}^{B^*}\vert^2}{2\eta_B}\cdot\one\{E_t^B\}\leq \sum_{s=1}^{S_b}\frac{\vert\mu_{w_s'}^{B}-\mu_{w_s'}^{B^*}\vert^2-\vert\mu_{e_s'+1}^{B}-\mu_{e_s'+1}^{B^*}\vert^2}{2\eta_B}.\label{eqn: main-reg-budget}
\end{align}
For $s\geq 2$, Note that we have $\mu_{w_s'-1}^{B}\leq \mu_{w_s'-1}^{B^*}$ and according to the update of $\mu_{t}^{B}$, we also have $\mu_{w_s'}^{B}\leq \mu_{w_s'-1}^{B}+\eta_B(\overlinev+\rho)$. Combining the fact that $\mu_{w_s'}^{B}\geq \mu_{w_s'}^{B^*}$, we have
\begin{align*}
    \vert\mu_{w_s'}^{B}-\mu_{w_s'}^{B^*}\vert^2 &\leq \vert\mu_{w_s'-1}^{B}+\eta_B(\overlinev+\rho)-\mu_{w_s'}^{B^*}\vert^2 \leq \vert\mu_{w_s'-1}^{B^*}+\eta_B(\overlinev+\rho)-\mu_{w_s'}^{B^*}\vert^2 \\
    &\leq  \left(\frac{\overlinev}{\rho}+2\eta_B(\overlinev+\rho)\right)\vert\mu_{w_s'-1}^{B^*}-\mu_{w_s'}^{B^*}\vert + \eta_B^2(\overlinev+\rho)^2.
\end{align*}
Combining the above inequality with \pref{eqn: main-reg-budget} finishes the proof.
$\hfill\qed$



\newpage
\section{Numerical Experiments: plots}
\label{app:expt-plots}
\input{{\PaperPath}expt-plots}

\end{document}